\documentclass{article}
\usepackage{arxiv}
\usepackage[pages=all, color=black, position={current page.south}, placement=bottom, scale=1, opacity=1, vshift=5mm]{background}

\usepackage{amssymb}
\usepackage{graphicx}
\usepackage{scalerel}
\usepackage{tikz, cite}

\usetikzlibrary{intersections}
\usetikzlibrary{shapes.geometric}
\usetikzlibrary{spy}
\usepackage{etoolbox}
\usepackage{pgfplots}
\usepackage{microtype}
\usepackage{fancyhdr}
\usepackage{multirow}
\usepackage{pgfplotstable}
\usepackage{amsmath,amssymb,amsfonts}
\usepackage{tabularx} 
\usepackage{adjustbox}
\usepackage{caption}
\usepackage{booktabs}
\usepackage{textcomp}
\usepackage{mathtools}
\usepackage{etoolbox}
\usepackage{color, colortbl,xcolor}
\usepackage{lineno,hyperref}
\usepgfplotslibrary{colorbrewer}
\usepackage[utf8]{inputenc} 
\usepackage[T1]{fontenc}
\usepackage[upright]{fourier} 
\usetikzlibrary{arrows}
\usepackage{tkz-kiviat,numprint} 

\usepgfplotslibrary{statistics}

\usetikzlibrary{positioning,shapes,arrows,shadows,patterns}
\mathchardef\mhyphen="2D
\newcommand\rnumber{\mathop{\mhyphen}}

\makeatletter
\def\tkzKiviatGrad{\pgfutil@ifnextchar[{\tkz@KiviatGrad}{\tkz@KiviatGrad[]}} 
\def\tkz@KiviatGrad[#1](#2){%
\begingroup
\pgfkeys{/kiviatgrad/.cd,
graduation distance= 0 pt,
prefix ={},
suffix={},
unity=1
 }
 \pgfqkeys{/kiviatgrad}{#1}%
\let\tikz@label@distance@tmp\tikz@label@distance
\global\let\tikz@label@distance\tkz@kiv@grad
 \foreach \nv in {0,...,\tkz@kiv@lattice}{ 
 \pgfmathparse{\tkz@kiv@unity*\nv} 
 \pgfmathtruncatemacro{\result}{\pgfmathresult-100} 
 \protected@edef\tkz@kiv@gd{\tkz@kiv@prefix$\result$\tkz@kiv@suffix}
    \path[/kiviatgrad/.cd,#1] (0:0)--(360/\tkz@kiv@radial*#2:\nv*\tkz@kiv@gap) 
       node[label=(360/\tkz@kiv@radial*#2)-90:\tkz@kiv@gd] {}; 
      }
 \let\tikz@label@distance\tikz@label@distance@tmp  
\endgroup
}%
\makeatother

\usetikzlibrary{svg.path}
\definecolor{orcidlogocol}{HTML}{A6CE39}
\tikzset{
  orcidlogo/.pic={
    \fill[orcidlogocol] svg{M256,128c0,70.7-57.3,128-128,128C57.3,256,0,198.7,0,128C0,57.3,57.3,0,128,0C198.7,0,256,57.3,256,128z};
    \fill[white] svg{M86.3,186.2H70.9V79.1h15.4v48.4V186.2z}
                 svg{M108.9,79.1h41.6c39.6,0,57,28.3,57,53.6c0,27.5-21.5,53.6-56.8,53.6h-41.8V79.1z M124.3,172.4h24.5c34.9,0,42.9-26.5,42.9-39.7c0-21.5-13.7-39.7-43.7-39.7h-23.7V172.4z}
                 svg{M88.7,56.8c0,5.5-4.5,10.1-10.1,10.1c-5.6,0-10.1-4.6-10.1-10.1c0-5.6,4.5-10.1,10.1-10.1C84.2,46.7,88.7,51.3,88.7,56.8z};
  }
}

\newcommand\orcidicon[1]{\href{https://orcid.org/#1}{\mbox{\scalerel*{
\begin{tikzpicture}[yscale=-1,transform shape]
\pic{orcidlogo};
\end{tikzpicture}
}{|}}}}

\newenvironment{customlegend}[1][]{%
        \begingroup
        \csname pgfplots@init@cleared@structures\endcsname
        \pgfplotsset{#1}%
    }{%
        \csname pgfplots@createlegend\endcsname
        \endgroup
    }%
    \def\addlegendimage{\csname pgfplots@addlegendimage\endcsname}

\pgfplotsset{compat=1.11,
    /pgfplots/ybar legend/.style={
    /pgfplots/legend image code/.code={%
       \draw[##1,/tikz/.cd,yshift=-0.25em]
        (0cm,0cm) rectangle (3pt,0.8em);},
   },
}

\begin{document}

\title{Auto-SpMV: Automated Optimizing SpMV Kernels on GPU}

\date{} 

\author{
        Mina~Ashoury \\
	Department of Computer Science and Engineering,          Shiraz University, Shiraz, Iran \\
	\texttt{minaashoury@gmail.com} \\
\AND
	Mohammad~Loni \\
        School of Innovation, Design and Engineering, M\"alardalen University, V\"aster\aa s, Sweden \\
	\texttt{mohammad.loni@mdu.se}
\AND
        Farshad~Khunjush \\
        Department of Computer Science and Engineering, Shiraz University, Shiraz, Iran \\
	\texttt{khunjush@shirazu.ac.ir}
 \AND
        Masoud~Daneshtalab \\
         School of Innovation, Design and Engineering, M\"alardalen University, V\"aster\aa s, Sweden \\
	\texttt{masoud.daneshtalab@mdu.se}
}

\maketitle

\begin{abstract}
Sparse matrix-vector multiplication (SpMV) is an essential linear algebra operation that dominates the computing cost in many scientific applications.
Due to providing massive parallelism and high memory bandwidth, GPUs are commonly used to accelerate SpMV kernels.
Prior studies mainly focused on reducing the latency of SpMV kernels on GPU.
However, few attempts have been made to improve the energy efficiency of SpMV kernels, resulting in GPUs being excluded from the range of low-power applications.
Furthermore, prior work has primarily focused on optimizing the sparse format of SpMV kernels, the literature ignores evaluating the impact of tweaking compilation parameters.
Lastly, Little attention has been paid to preparing a comprehensive training dataset of running SpMV kernels and fine-tuning the learning hyperparameters.
To address these limitations, we present a novel framework, dubbed Auto-SpMV, that enables energy-efficient and low-latency SpMV kernels on GPU.
To achieve the best run time performance, Auto-SpMV proposes two optimization modes: \textit{compile-time} and \textit{run-time}.
In the \textit{compile-time} mode, Auto-SpMV tweaks the compilation parameters, while in the \textit{run-time} mode, Auto-SpMV selects the best sparse format for the input matrix.
To achieve the best classification results, 1) we collect the largest dataset ever having 30 different sparse matrices running with more than 15K different configurations, and 2) we boost classification models by automatically fine-tuning the learning hyperparameters. 
Experimental results reveal that Auto-SpMV optimizes latency, energy consumption, average power, and energy efficiency in the \textit{compile-time} mode by up to 51.9\%, 52\%, 33.2\%, and 53\%, respectively, compared to the default setting.
Auto-SpMV optimizes average power and energy efficiency in the \textit{run-time} mode by up to 34.6\% and 99.7\%, respectively, compared to the default setting. 

\end{abstract}


\section{Introduction}
\label{sec:Introduction}

Sparse matrix-vector multiplication (SpMV) is a key operation in a variety of scientific applications such as large simulation systems \cite{oyarzun2021fpga}, graph learning \cite{qiu2021optimizing}, economic modeling \cite{schiesser2014computational}, and more.
It is extremely imperative to optimize the SpMV kernel on modern hardware devices because SpMV dominates the computing cost in iterative problem-solving methods such as the calculation of Eigenvalues \cite{sgherzi2022mixed} and Krylov subspace \cite{ahamed2014accelerated}.
GPUs are high-throughput devices that are widely used for accelerating a variety of scientific applications due to the ease of programming and the ability to provide large amounts of processing power and memory bandwidth.

Prior studies investigated how to improve the performance of SpMV kernels on GPU.
\cite{guo2011model,zardoshti2016adaptive,abu2012effective} proposed automated tweaking of different parameters, such as the maximum number of registers per thread or thread block size according to the sparsity pattern of the input matrix.
Another research direction was related to reducing the memory footprint of SpMV kernels by proposing different efficient sparse formats such as COO \cite{bell2009implementing}, CSR \cite{bell2009implementing}, ELL \cite{bell2009implementing}, BELL \cite{choi2010model}, and SELL \cite{choi2010model}.
However, as shown in many studies \cite{[31], filippone2017sparse, benatia2016machine}, using an inappropriate sparse format can result in significant performance degradation (up to 1.6$\times$ \cite{lehnert2016performance}).
To address this limitation, recent research studies proposed selecting the optimal sparse format automatically to minimize the execution time and/or reduce the energy consumption of SpMV kernels on GPU \cite{[15], [16], zhao2018bridging, [31]}.
The proposed sparse format selection methods have been successful, but they ignore the importance of optimizing power consumption (MFLOPS/Watt), which prevents GPUs from processing SpMV kernels in many low-power applications \cite{mach2017mobile}. 
Furthermore, GPU performance of the SpMV kernel has a high sensitivity to the compiler configuration parameters (thread block size, maximum number of registers per thread, and memory hierarchy configuration) \cite{zardoshti2016adaptive}; nonetheless, the proper selection of these parameters has not been explored in the literature.  
Finally, it is not conclusive that the results of the prior works are reproducible, primarily since a comprehensive evaluation cannot be found in their report \cite{lindauer2020best}.

To tackle shortcomings of the state-of-the-art methods, we propose a machine learning framework that automatically tweaks both sparse format and compilation parameters, targeting different optimization objectives such as 1) latency (Second), 2) energy consumption (Joule), 3) average power consumption (Watt) and 4) energy efficiency (MFLOPS/Watt) of the SpMV kernel on GPU.
We call this method automated tweaking of the SpMV kernel configuration parameters or Auto-SpMV.
Auto-SpMV enables two modes of optimization: \textit{compile-time} and \textit{run-time}. In the \textit{compile-time} mode, Auto-SpMV tweaks compiler parameters while in the \textit{run-time} mode, Auto-SpMV automatically selects the most efficient sparse format and programming kernel according to the structural characteristics of the sparse input matrix.

In this paper, we consider CSR, ELL, BELL, and SELL sparse formats due to simple implementation and providing higher performance in many applications over other hybrid formats \cite{filippone2017sparse}.
To characterize the performance of SpMV kernels for sparse matrices, we consider simple sparsity features of the sparse input matrix. 
In the learning process, a multi-class classifier is trained on a dataset of sparse matrices characterized by their sparsity features.
Then, we use this model to predict the optimal sparse format, maximum number register per thread (\texttt{maxrregcount}), memory hierarchy configuration, and thread block size (TB size) for unseen inputs.

To achieve maximum learning performance, the classifier should be trained on a comprehensive dataset. 
In order to fulfill this requirement, we collect a dataset of 30 sparse matrices from the SuiteSparse matrix collection \cite{davis2011university} that contains records of SpMV kernels running on GPU obtained from different compiler parameters and sparse formats.
To the best of our knowledge, Auto-SpMV provides the largest training dataset containing 15520 records with $\approx$70 runs on two GPUs (Turing and Pascal architectures) compared to all existing studies on automated SpMV optimization.
To improve the classification accuracy, we need to fine-tune the learning hyperparameters such as the learning rate of neural networks. 
In order to fulfill this requirement, we use an Automated Machine Learning (AutoML) tool to efficiently optimize the learning pipeline (Section~\ref{sec:Method:ML}). 
Then, we report the best classification results.  
In this paper, we also devise regression models to estimate various optimization objectives on a sparse input matrix.

Our main contributions are summarized as follows:

\begin{enumerate}
    \item Enhancing the performance of SpMV kernels on GPU through the use of efficient machine learning classifiers. This will enable us to select the optimal compilation parameters and sparse format based on different optimization objectives.
    
    \item As the general demand for green GPU processing grows \cite{qasaimeh2019comparing}, Auto-SpMV considers new optimization objectives: energy efficiency and average power consumption. Along with latency and energy consumption, evaluating these two optimization objectives is extremely tedious since an SpMV kernel must be run many times on GPU.

    \item Using highly-optimized classifiers rather than relying on widely-used models trained with default learning hyperparameters.
    
    \item Collecting the largest ever dataset containing the record of running 30 different sparse matrices with more $\approx$70M runs over 15k configuration settings on two different GPU architectures. Dataset, code, and configuration parameters will be available upon acceptance.
\end{enumerate}

Results show that Auto-SpMV optimizes latency, energy consumption, average power, and energy efficiency in the \textit{compile-time} mode by up to 51.9\%, 52\%, 33.2\%, and 53\%, respectively, over all selected sparse matrices compared to the best default configuration settings.
Auto-SpMV optimizes average power and energy efficiency in the \textit{run-time} mode by up to 34.6\%, and 99.7\%, respectively.
We analyze the overhead introduced by the run time kernel selection and translation between different sparse formats.
Finally, Our experimental results on two different NVIDIA\textsuperscript{®} GPU devices, GTX 1650-mobile (Turing architecture) and GTX 1080 (Pascal architecture), show that Auto-SpMV prediction results are not sensitive to the underneath GPU architecture.
%
%

The rest of this paper is organized as follows: 
Section~\ref{sec:Background} introduces the sparse formats considered and how the SpMV kernel is executed on GPU. 
In Section~\ref{sec:Motivation}, we introduce the problem of selecting sparse formats and compiler parameters.
In Section~\ref{sec:Method}, we present the Auto-SpMV optimization framework.
Section~\ref{sec:Configuration} presents the configuration setup for evaluating our proposed method.
In Section~\ref{sec:Results}, we evaluate the Auto-SpMV framework and compare it to the other baselines. 
We thoroughly discuss the proposed results and analyze the processing overhead of Auto-SpMV in Section~\ref{sec:Discussion}. 
In Section~\ref{sec:RelatedWork}, we review the related studies.
Finally, we conclude our paper in Section~\ref{sec:Conclusion}.

\section{Background}
\label{sec:Background}

\subsection{GPU Architecture}
\label{sec:Background:GPU_Architecture}

GPUs are designed to maximize computing throughput by executing parallel workloads, even though sacrificing a single task's serial performance.
GPUs offer a large number of Streaming Processors (SPs) equipped with fully-pipelined logic units.
SPs are grouped in a set of Streaming Multiprocessors (SMs) to execute instructions in Single-Instruction Multiple-Threads (SIMT) mode.
A kernel is a GPU-related part of an application and is executed by thousands of threads arranged in a grid of thread blocks defined by the programmer.
Thread blocks are the groups of threads that are executed simultaneously in SMs.
The SM runs threads in groups of 32 parallel threads called warps.
To achieve the maximum throughput, every SM unit should have enough active warps to hide memory and instruction pipeline latencies. 

\begin{figure}[htbp]
\begin{center}
\centerline{\includegraphics[width=0.75\columnwidth]{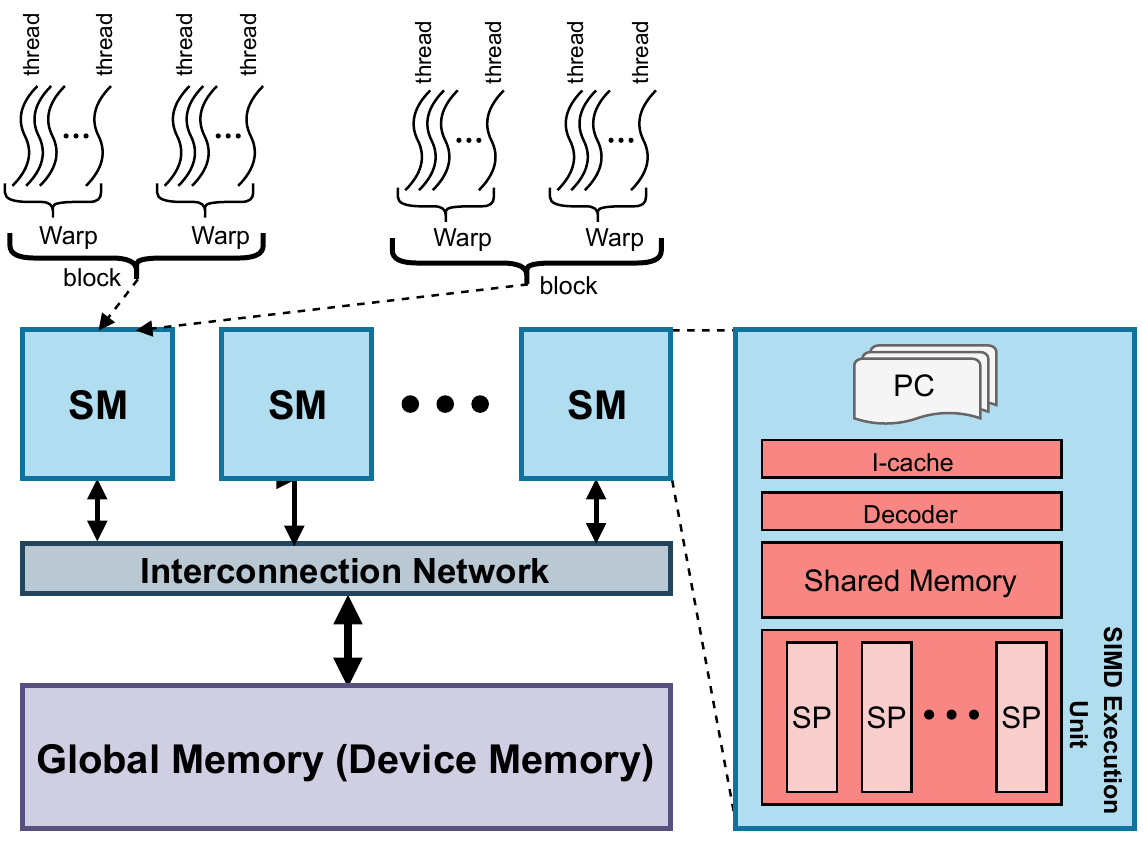}}
\caption{An overview of GPU architecture.}
\label{fig:GPU_Arch}
\end{center}
\end{figure}

GPUs feature a rich memory hierarchy configuration that includes L1 and L2 caching capabilities, texture memory, shared memory, and global memory.
Memory hierarchy can be defined by the programmer, allowing L1 cache and shared memory to be configured manually. 
Thread blocks use shared memory that can be accessed and shared by all threads within a block.
Programmers can identify the upper-bound number of registers used by a thread at the compile time by using the \texttt{maxrregcount} variable.
According to the default settings, CUDA\textsuperscript{®} compilers try to minimize register usage to maximize the number of thread blocks that can be active simultaneously within the SM, leading to maximizing occupancy within the SM.
An overview of GPU architecture is illustrated in Figure~\ref{fig:GPU_Arch}.

\subsection{SpMV Processing on GPU}
\label{sec:Background:SpMV_Processing_GPU}

The Sparse Matrix-Vector Multiplication (SpMV) is a non-trivial Level-2 BLAS operation that finds the dense vector product $Y$ of a sparse matrix $A$ and a dense vector $X$ such that $Y = A \times X$.
Figure~\ref{Fig:SpMV_Formats}.a shows an example of SpMV operation for matrix $A$ and dense vector $X$.
Dense formats are inefficient for implementing SpMV kernels on GPU due to storing zeros and the zero multiplications result in inefficient memory usage and computation, respectively.
Several sparse formats have been proposed to address this problem (Section~\ref{sec:Background:SpMV_Formats}). 

\subsection{Popular SpMV Formats}
\label{sec:Background:SpMV_Formats}

Several sparse formats have been proposed in order to improve the efficiency of running SpMV kernels on GPU by avoiding unnecessarily 1) storing zero elements, and 2) performing computations on them. 
Note that there is no best-for-all sparse format for all SpMV matrices \cite{[31], filippone2017sparse, benatia2016machine}.
In other words, if a sparse format is not carefully selected, significant performance degradation is likely to occur \cite{[31]}.
Due to the sparsity features of matrices, the sparse format can be used in flat, blocked, and composite types. 
Here is a brief explanation of the most popular sparse formats that we used in this paper.

\textbf{CSR format \cite{bell2009implementing}.} According to the CSR format, non-zero elements and column indices are stored explicitly in \emph{Data} and \emph{Column Index} arrays, respectively. The boundaries of each row are saved in a third array called \emph{Row Index}. Figure~\ref{Fig:SpMV_Formats}.b shows the CSR format of matrix $A$. This format requires coordination among threads within a warp to accumulate per-thread results together. Although this format requires no zero-padding, the access to the \emph{Data} and \emph{column Index} might be unaligned. Having random access to $X$, as well as the varied number of non-zero elements per row, this kernel may suffer from the load imbalance problem.

\textbf{ELL format \cite{bell2009implementing}.} In the ELL format, non-zero elements are grouped into a dense \emph{Data} matrix, and the column indices of matrix $A$ are stored in a \emph{Column Index} matrix. Dimensions of the \emph{Data} matrix will be $m \times max\_nnz$, in which $max\_nnz$ is the maximum number of non-zero elements in a row. Figure~\ref{Fig:SpMV_Formats}.c shows the ELL format of matrix $A$. Despite the efficiency of the ELL format, it does not necessarily guarantee contiguous access to the input vector $X$. 

\textbf{Blocked ELL (BELL) format \cite{choi2010model}.} This is a variation of the ELL format in which a block of non-zero elements is considered as an element of the ELL format. Figure~\ref{Fig:SpMV_Formats}.d shows the BELL format of matrix $A$ with the default block size of 2$\times$2. The BELL format is composed of two matrices: 1) \emph{Data} matrix, which is used to store blocks of non-zero elements, and 2) \emph{Column Index} matrix, which is used to store the block indices. In general, this format is suitable for matrices with a uniform distribution of non-zero blocks.

\textbf{Sliced ELL (SELL) format \cite{choi2010model}.} Each slice of this format consists of a constant number of rows of matrix $A$. The length of each slice is the maximum number of non-zero elements per row in that slice. For compressing a matrix to SELL format, each slice of the matrix must be considered as a matrix packing in ELL format, and this is achieved by storing the pointers of each slice. SELL format is composed of three matrices including 1) \emph{Slice Index} saves pointers to slices; 2) \emph{Data} matrix stores slices of non-zero elements; and 3) \emph{Column Index} matrix stores column indices for the \emph{Data} elements. Figure~\ref{Fig:SpMV_Formats}.e shows the SELL format of matrix $A$ with slice height set to 2. This format is suitable for sparse matrices with a variety of elements per row that are non-zero. 

\begin{table*}[htbp]
\centering
\captionsetup{justification=centering}
\resizebox{\textwidth}{!}{
\begin{tabular}{ccccc}

\includegraphics[width=0.26\textwidth]{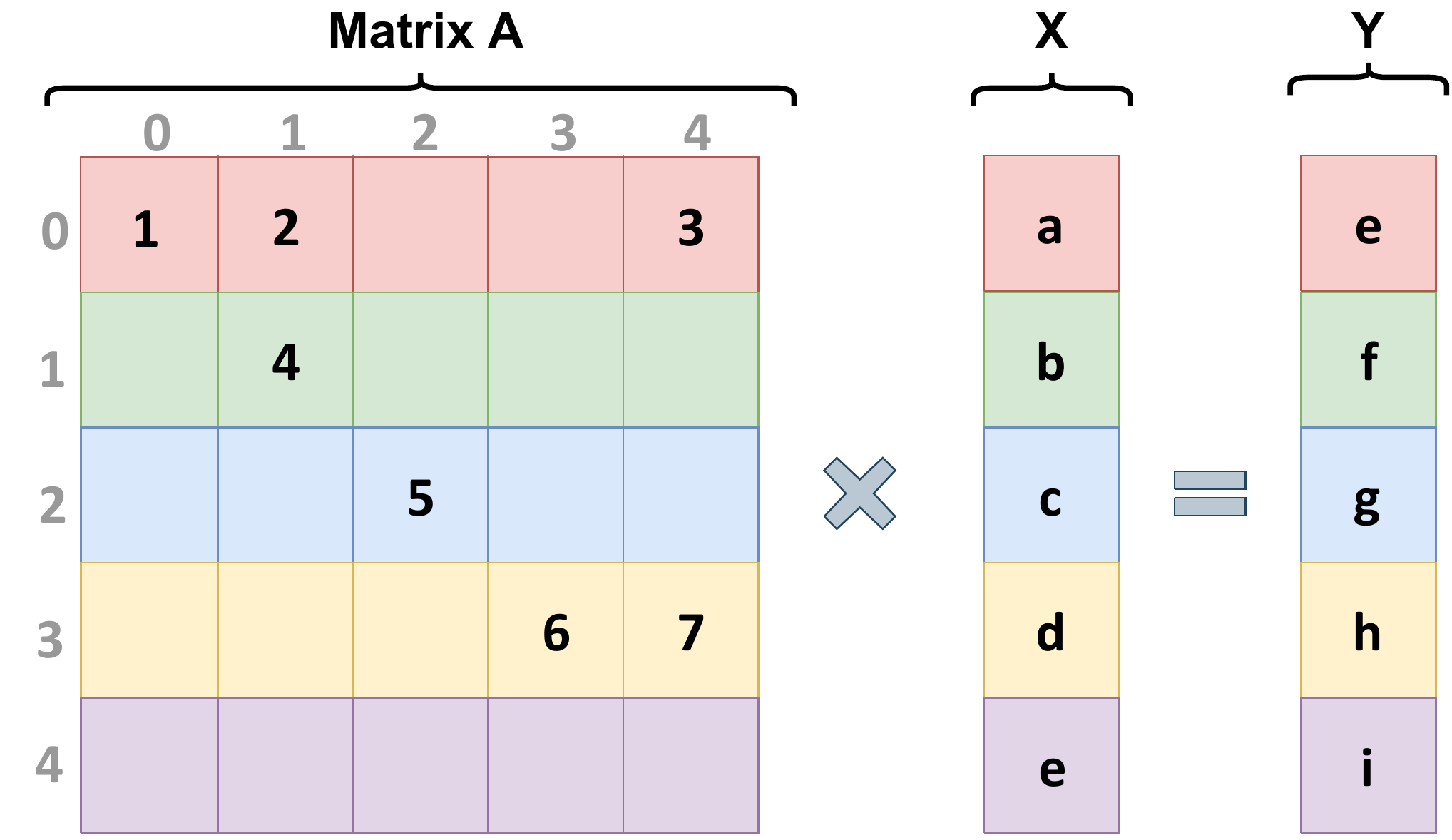}&
\includegraphics[width=0.24\textwidth]{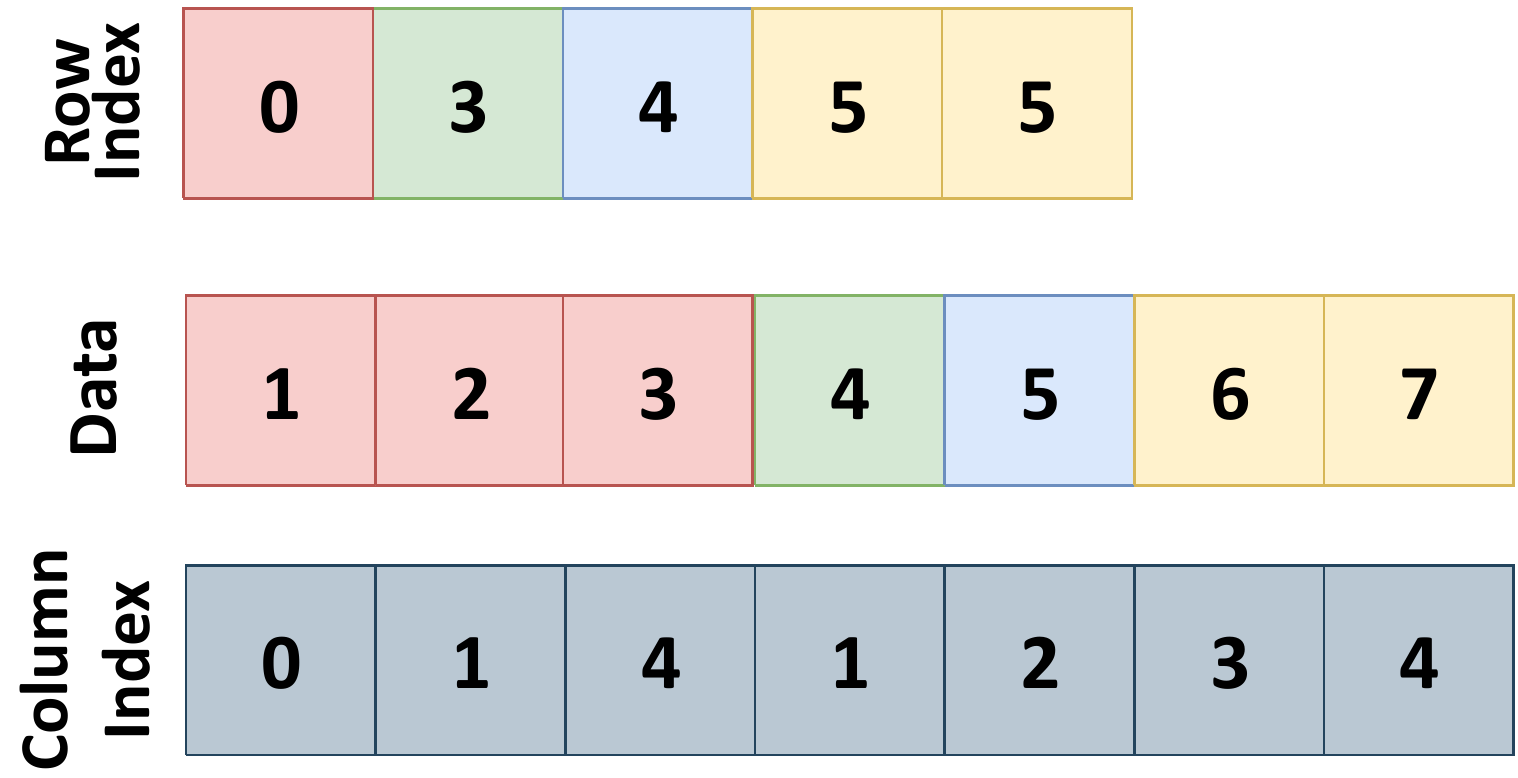}&
\includegraphics[width=0.2\textwidth]{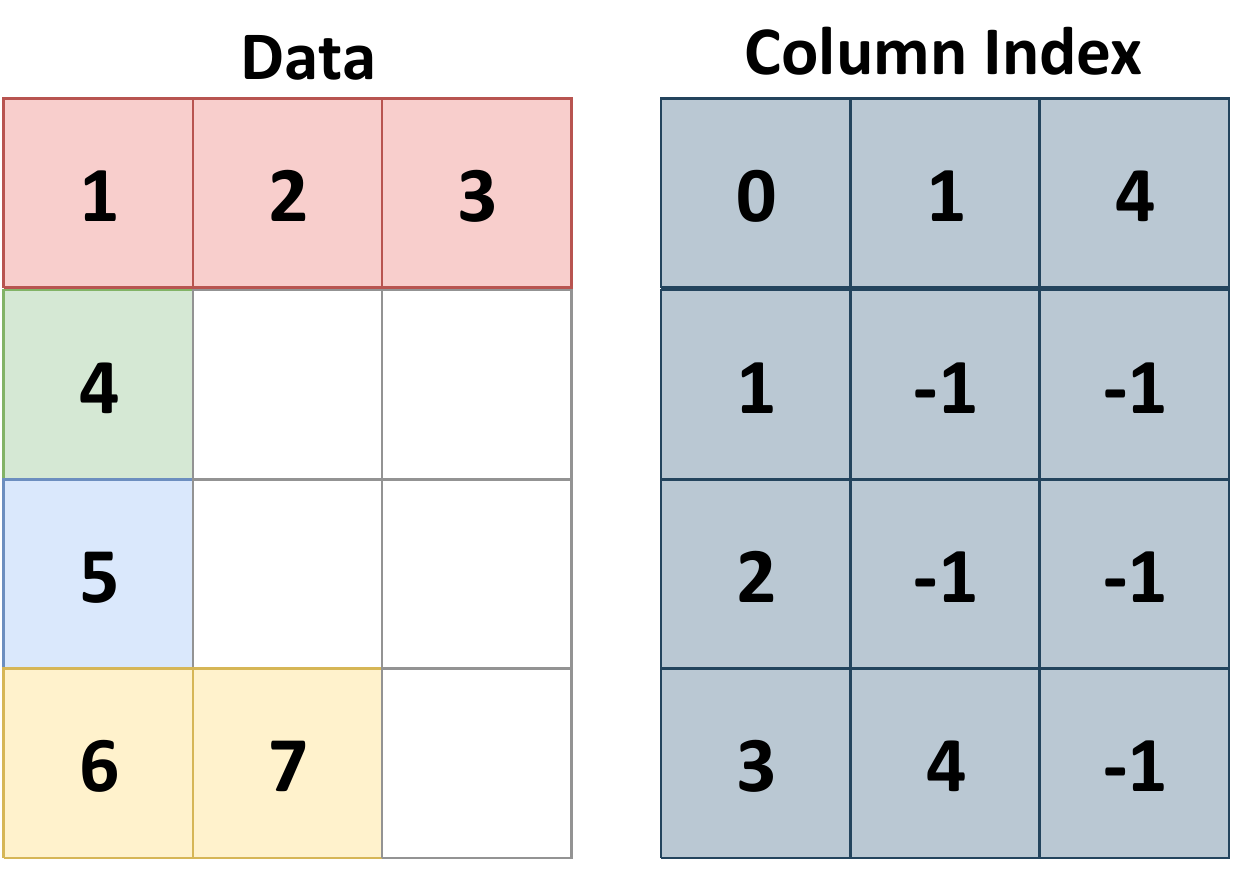}&
\includegraphics[width=0.21\textwidth]{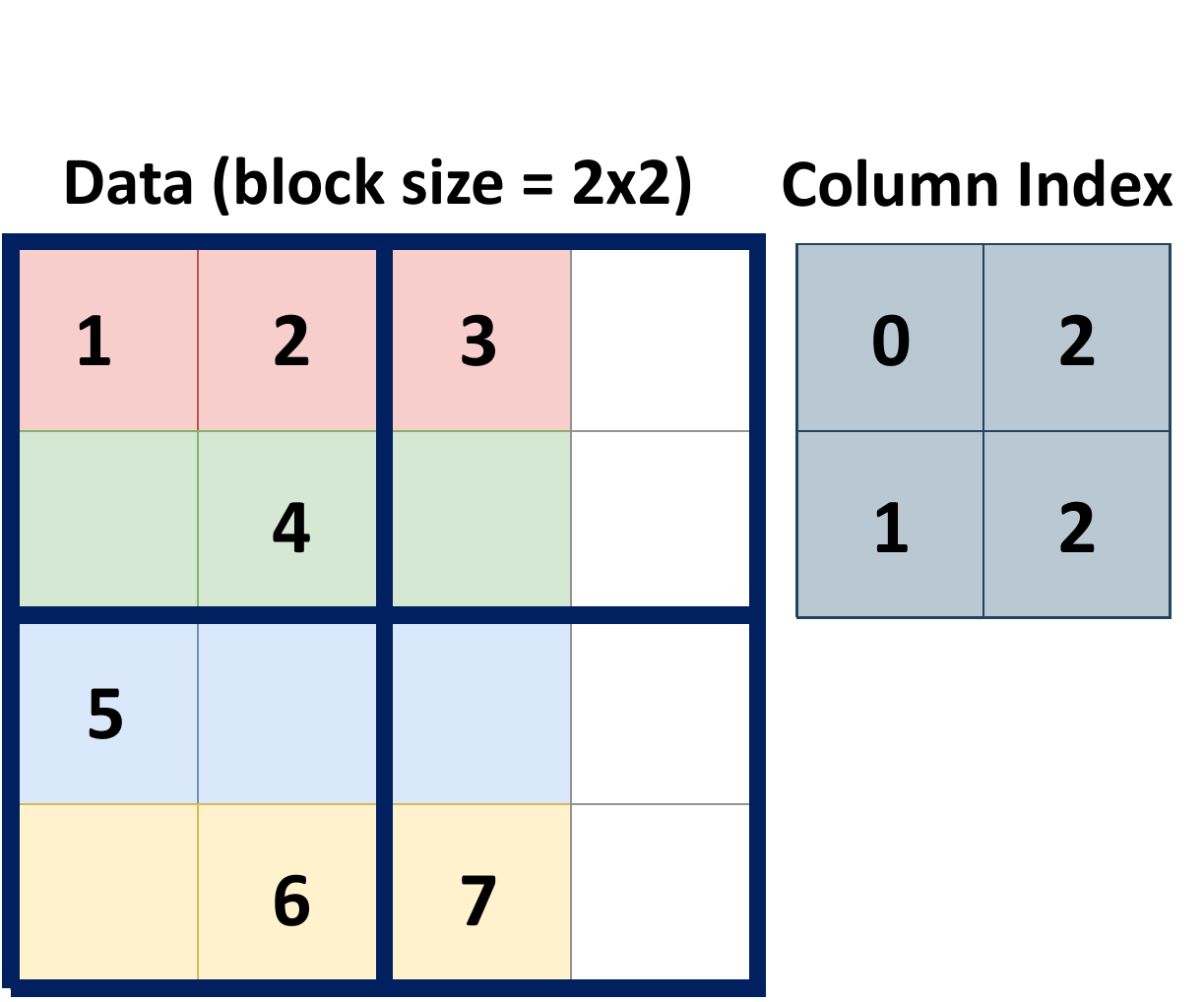}&
\includegraphics[width=0.22\textwidth]{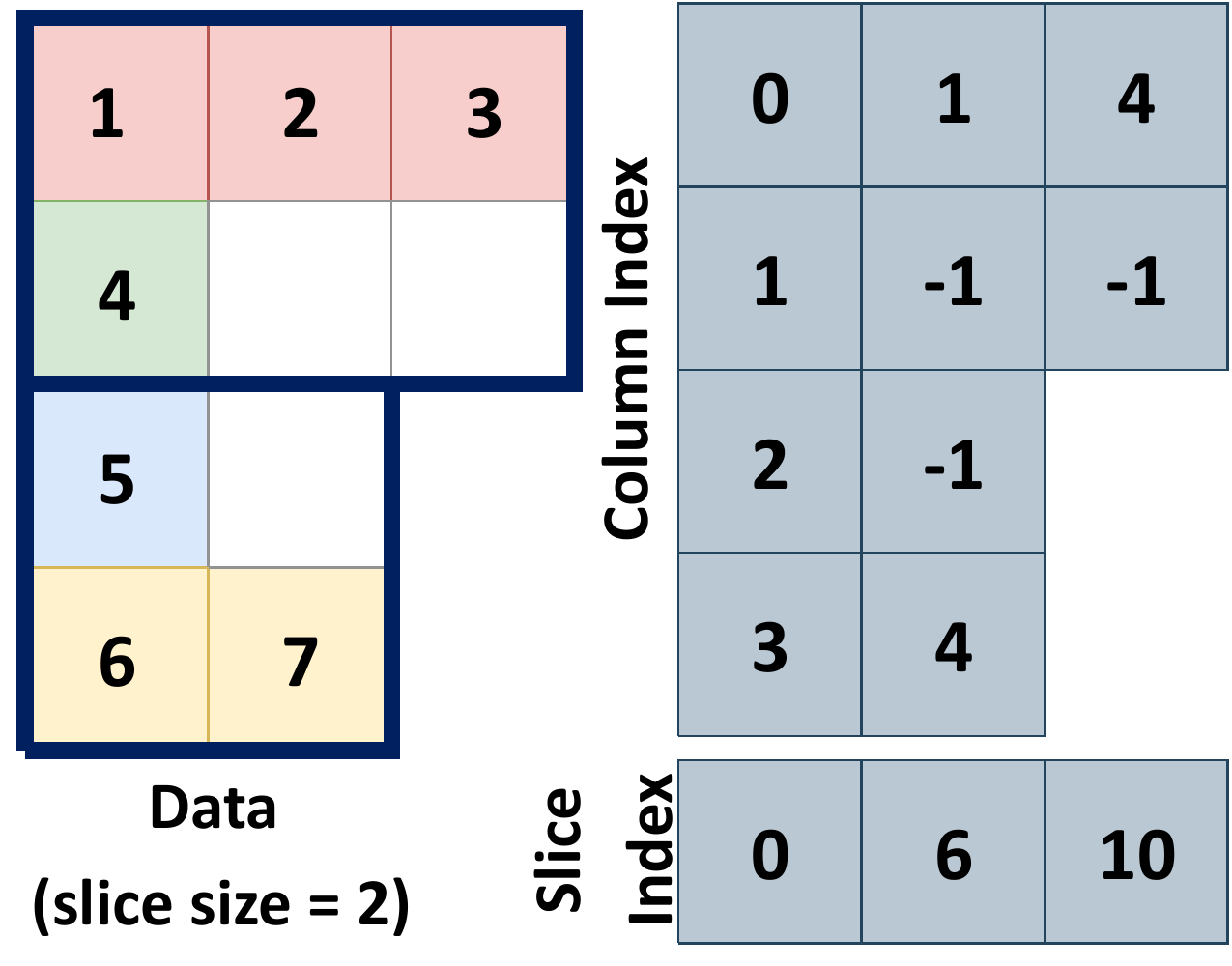}
\\
\textbf{(a) Input Matrix} & \textbf{(b) CSR} & \textbf{(c) ELL} & \textbf{(d) BELL} & \textbf{(e) SELL}
\\

\end{tabular}}
\captionof{figure}{Illustration of (a) sparse matrix-vector multiplication, (b) CSR, (c) ELL, (d) BELL, and (e) SELL formats. Empty colorful cells are zero values of the sparse input matrix and white cells are zero padding.}
  \label{Fig:SpMV_Formats}
\end{table*}

\section{Research Motivation}
\label{sec:Motivation}

In this section, we first demonstrate the importance of tweaking compilation time parameters and the inefficiency of the default optimization scheme.
Then, we discuss the limitations of prior learning models that attempted to model the latency and energy consumption of an SpMV kernel on GPU.

\subsection{Inefficiency of the Default Optimization Parameters}
\label{sec:Motivation:Inefficiency}

Figure~\ref{fig1:Motivation_1} compares the results of optimizing storage format and compiler parameters of the SpMV kernel by Auto-SpMV with the default CUDA\textsuperscript{®} compiler configuration parameters for the \textit{consph} sparse matrix given as a sample benchmark.  
Results are normalized to the Auto-SpMV (the higher is better).
Because the CSR \cite{bell2009implementing} format is popular, we have selected it as CUDA's default sparse format. 
The results demonstrate that Auto-SpMV provides at least 2.04$\times$ lower latency, 2.07$\times$ lower energy consumption, 1.08$\times$ lower average power consumption, and 2.086\% better energy efficiency for the selected sparse matrix compared to the default CUDA\textsuperscript{®} compilation parameters.
Clearly, the default configuration parameters are not optimal, resulting in a significant degradation for all optimization objectives.
Optimizing the compiler configuration parameters and the sparse format of SpMV kernels is therefore of paramount importance.

\begin{table}[htbp]
\centering
\resizebox{\columnwidth}{!}{
\begin{tabular}{c}

\begin{tikzpicture}
\begin{customlegend}[legend columns=5,legend style={text opacity = 1,row sep=0pt, font=\fontsize{40}{8}\selectfont, column sep=2ex},
        legend entries={{Latency},
                        {Energy~Cons.},
                        {Avg.~Power},
                        {Energy~Eff.},
                        {Optimal (Auto-SpMV)}
                        }]
        \addlegendimage{mark=square*, mark size=11pt, only marks, red, thick, draw=black}
        \addlegendimage{mark=square*, mark size=11pt, only marks, teal, thick, draw=black}
        \addlegendimage{mark=square*, mark size=11pt, only marks, orange, thick, draw=black}
        \addlegendimage{mark=square*, mark size=11pt, only marks, violet, thick, draw=black}
        \addlegendimage{mark=square*, mark size=11pt, only marks, gray, thick, draw=black}
        \end{customlegend}
\end{tikzpicture}
\\
\begin{tikzpicture}
\begin{axis}[
    ybar=12pt,
    width=3\columnwidth,
    height=0.75\columnwidth,
    font=\Huge,
    enlargelimits=0.15,
    grid=major, 
    grid style={dashed,gray!90}, 
    legend style={at={(0.5,1.15)}, nodes={scale=1},
      anchor=north,legend columns=-1},
    ylabel={\Huge \textbf{Normalized Value}},
    xlabel={\Huge \textbf{Thread Block Size}},
    symbolic x coords={TB=128,TB=256,TB=512,TB=1024},
    xtick=data,
    ytick={0.4,0.7,1},
    bar width=28pt,
    nodes near coords align={vertical},
    ]
\addplot [red, fill, draw=black ] coordinates {(TB=128,0.487545071) (TB=256,0.48929839) (TB=512,0.489791284) (TB=1024,0.480528473)};

\addplot [teal, fill, draw=black] coordinates {(TB=128,0.479190177) (TB=256,0.482088793) (TB=512,0.48159727) (TB=1024,0.47163974)};

\addplot  [orange, fill, draw=black] coordinates {(TB=128,0.924127568) (TB=256,0.922478325) (TB=512,0.921072894) (TB=1024,0.920024794)};

\addplot [violet, fill, draw=black] coordinates {(TB=128,0.478625374) (TB=256,0.479489367) (TB=512,0.479241123) (TB=1024,0.469642813)};

\addplot [gray , fill, draw=black] coordinates {(TB=128,1) (TB=256,1) (TB=512,1) (TB=1024,1)};

\end{axis}
\end{tikzpicture}

\end{tabular}
}

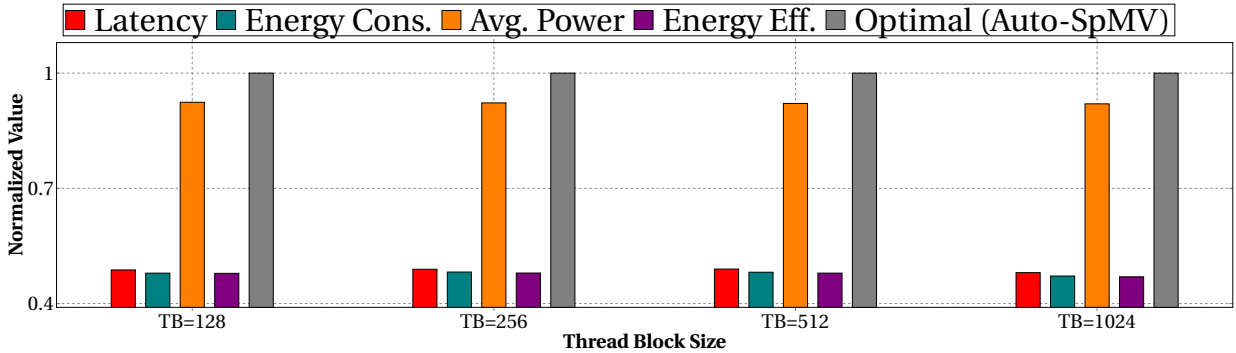
\captionof{figure}{Comparing Auto-SpMV results with the default configuration parameters for the \textit{consph} sparse matrix. Auto-SpMV is selected as the normalization baseline (the higher is better).}
\label{fig1:Motivation_1}
\end{table}

\subsection{Limitations of Existing Learning Models}
\label{sec:Motivation:SpMV_Tweaking_challenges}

Due to the complexity of GPU architectures and the wide range of different sparsity features, it is very hard to describe the effect of each one of these sparsity features on the performance of SpMV kernels on GPU \cite{[31]}.
The lack of such thorough understanding is behind the motivation of using a machine learning approach.
Efficient execution of the SpMV kernels on GPU requires optimizing several software and hardware parameters, resulting in a large optimization space for selecting the optimal values.
Therefore, utilizing an exhaustive optimization algorithm to find the optimal solution is not feasible in practice since the procedure is extremely time-consuming \cite{loni2020deepmaker}.
Several analytical models have been proposed to estimate the performance of SpMV kernels and select the best sparse format \cite{guo2013performance}.
However, analytical models cannot cope with the complexity of modern applications due to hardware complexities and requiring assumptions to simplify the problem \cite{benatia2016machine}.

To tackle the existing challenges, prior studies \cite{benatia2016machine, [30], zhao2018bridging, [13], [14], [15]} proposed to devise machine learning models to 1) estimate the performance and energy consumption of SpMV kernels on GPU, and 2) predict the near-optimal configuration parameters. 
The use of machine learning models offers a number of benefits, including that they are independent of hardware specifications, such as the number of cores per SM.
Further, assumptions that simplify the problem are not necessary.
Despite the success of prior studies in building machine learning models, they suffer from the inaccurate prediction (up to 16\% accuracy loss \cite{[31]}) due to 1) collecting a relatively small training dataset, and 2) evaluating a limited number of machine learning models.
To address these challenges, Auto-SpMV devises highly accurate machine learning models trained on a large dataset consisting of 15520 samples.
Approximately 70 million runs were performed on two different NVIDIA\textsuperscript{®} GPU architectures, Pascal and Turing, in order to construct such a comprehensive training dataset.

\section{Configuration Parameters}
\label{sec:Configuration}

\begin{figure}[htbp]
    \centering
    \pgfmathsetmacro{\BarOffset}{0.15}
    \pgfplotsset{
    MyAxis/.style={
            ybar,
            scale only axis,
            width=0.85\columnwidth,
            height=2.5cm,
            ymin=0,
            ymax=15,
            xmin=-0.5,
            xmax=15.5, 
            area legend,
            bar width=10pt
      }
    }
\pgfplotstableread{
0 3.19 
1 1.7 
2 0 
3 3.56 
4 4.55	
5 4.033	
6 0	
7 4.69	
8 6.029 
9 5.981 
10 0.362 
11 4.82 
12 0 
13 0 
14 13.76 
15 0 
}\data
    \begin{tikzpicture}
    \begin{axis}[
        ylabel={\small \textbf{Improvement (\%)}},
        ymin=0,
        ymax=15,
        xtick={0,...,15},
        xtick pos=left,
        ytick pos=left,
        font=\scriptsize,
        xticklabels = {Energy~Cons., Latency, Avg.~Power, Energy~Eff., Energy~Cons., Latency, Avg.~Power, Energy~Eff., Energy~Cons., Latency, Avg.~Power, Energy~Eff., Energy~Cons., Latency, Avg.~Power, Energy~Eff.},
        xticklabel style={rotate=90},
        yticklabel style={xshift=-0.5ex},
        tickwidth=5pt,
        extra x ticks={-0.5,3.5,7.5,...,15.5},
        extra x tick labels={},
        tickwidth=0mm,
        ymajorgrids,
        every node near coord/.append style={anchor=north},
        xtick style={draw=black},
        extra x tick style={tickwidth=0.9cm},
        minor x tick style = {opacity=1},
        MyAxis]
    \addplot[draw=black,fill=red] table[x expr=\coordindex-\BarOffset,y index=1] \data; 
    \label{returnplot}
    \end{axis}
    \begin{axis}[
        name=ax2,
        ymax=14,
        MyAxis,
        font=\scriptsize,
        yticklabel=\empty,
        yticklabel style={xshift=0.5ex},
        clip=false,
        xtick={1.5,5.5,9.5,13.5},
        xticklabels={\textbf{\texttt{maxrregcount}}, \textbf{TB Size}, \textbf{Memory  Conf.}, \textbf{ Sparse Format} },
        xticklabel style={yshift=-15mm},
        tickwidth=5pt,
        xtick style={draw=none}
    ]
    \end{axis}
    \end{tikzpicture}
    \caption{Individual improvement of each configuration parameter provided by Auto-SpMV over different optimization objectives. We select \textit{eu-2005} as the sample benchmark. }
    \label{fig1:Configuration}
    \end{figure}
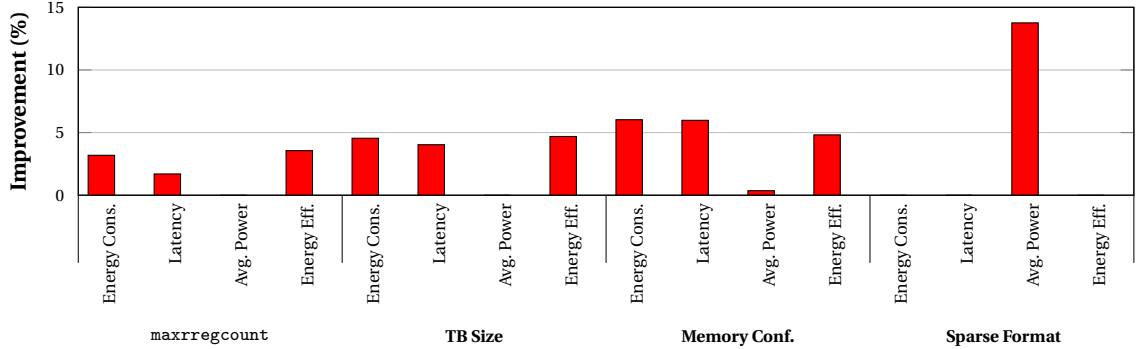

In this section, we provide an ablation study to individually examine the contribution of each compilation parameter and sparse format to the optimization objectives.
Figure~\ref{fig1:Configuration} shows the impact of optimizing each individual configuration setting of SpMV kernels by Auto-SpMV on various optimization objectives for the \textit{eu-2005} sparse matrix given as a sample benchmark.

The experimental results provided in Figure~\ref{fig1:Configuration} have led to the following observations:

\begin{enumerate}

\item \textbf{Maximum number of registers per thread (\texttt{maxrregcount}).} Register memory is the thread-private memory that is divided among all resident threads on an SM.
All local variables are placed in registers or local memory. Because local memory is placed in device DRAM (although it can be cached on-chip) and registers are on-chip, it is preferable for thread-private variables to be placed in registers.
The maximum number of registers that can be used per thread in a kernel is controlled by the compiler.
Nonetheless, the programmer can declare the number of registers used in every kernel in a compilation unit.
Limiting the number of registers per thread can lead to increasing the number of blocks that can concurrently reside on an SM, which by itself can result in better latency hiding. Nevertheless, restricting the number of registers can cause register spilling which occurs when there are not enough registers available for a given task.
Hence, registers can spill to global memory.
Because of the opposing factors of register spilling and higher occupancy, some experiments are often needed to obtain the optimal configuration \cite{fatica2014cuda}. 

\item \textbf{Thread block size (TB Size).} The SpMV kernel is a memory-bound application, and a kernel with a higher occupancy rate has more latency-hiding ability. For improving kernel occupancy, we need to increase the size of the thread block. However, the higher value for thread block size might result in imbalanced resource usage in each SM. Determining the proper thread block size while sharing the workload equally across all threads results in fewer idle threads, thereby leading to improved performance. On the other hand, occupancy is an indicator of thread parallelism in a CUDA\textsuperscript{®} program and can be increased by increasing the thread block size. However, maximizing the thread block size can lead to wasting achievable parallelism, e.g., in the case of suspended thread blocks, there are fewer other thread blocks to be active. Hence, the optimal number of threads per block is the result of the occupancy and performance trade-off and can be obtained after some experimentation.

\item \textbf{Memory hierarchy configuration.} For memory-bound kernels, such as SpMV, the optimization challenge is to reduce the memory latency time by leveraging the fast memory resources and caching capabilities of GPU. Exploiting the reconfigurability of the memory hierarchy and other hardware resources, such as registers, may lead to the energy optimization of GPU devices.  The ideal case is that after each thread’s first load, the next elements will reside in the L1 cache, but up to 2048 threads can run on an SM sharing L1 cache and likely replace each other’s data. Although increasing the size of the L1 cache on GPU for data-intensive applications is beneficial, there might be some additional latency when using L1 due to the need to translate the global address you are accessing to a cache location. Hence, various input matrices with a variable number of non-zero elements in rows have different cache and shared memory needs.

\item \textbf{Sparse format.} The irregular computations involved in SpMV make its performance optimization challenging.
Hence, enormous attempts have been devoted to devising data formats to store the sparse matrix with the aim of maximizing performance.
Due to the sparsity features of matrices, sparse formats can be used in flat, blocked, and composite types.
Therefore, we select four widely-used sparse formats which have fewer specifications to sparsity features and can be used generally.

\end{enumerate}

Although the main method for improving the efficiency of SpMV kernels on GPU is choosing the best sparse format \cite{kanellopoulos2019smash}; the kernel optimization does not end at this point. 
As demonstrated in this section, optimizing the compiler parameters  has a higher impact on improving the efficiency of SpMV kernels on GPU. 
Therefore, Auto-SpMV enables optimizing both \textit{compile-time} (thread block size, \texttt{maxrregcount}, and memory hierarchy configuration) and \textit{run-time} parameters (sparse format) by devising highly accurate machine learning models.
Furthermore, Auto-SpMV is capable of finding optimal configuration settings for a wider range of optimization objectives, including latency, energy consumption, energy efficiency, and average power consumption.
 
\section{Auto-SpMV Optimization Method}
\label{sec:Method}

\subsection{Method Overview}
\label{sec:Method:Overview}

\begin{figure}[ht]
\begin{center}
\centerline{\includegraphics[width=\columnwidth]{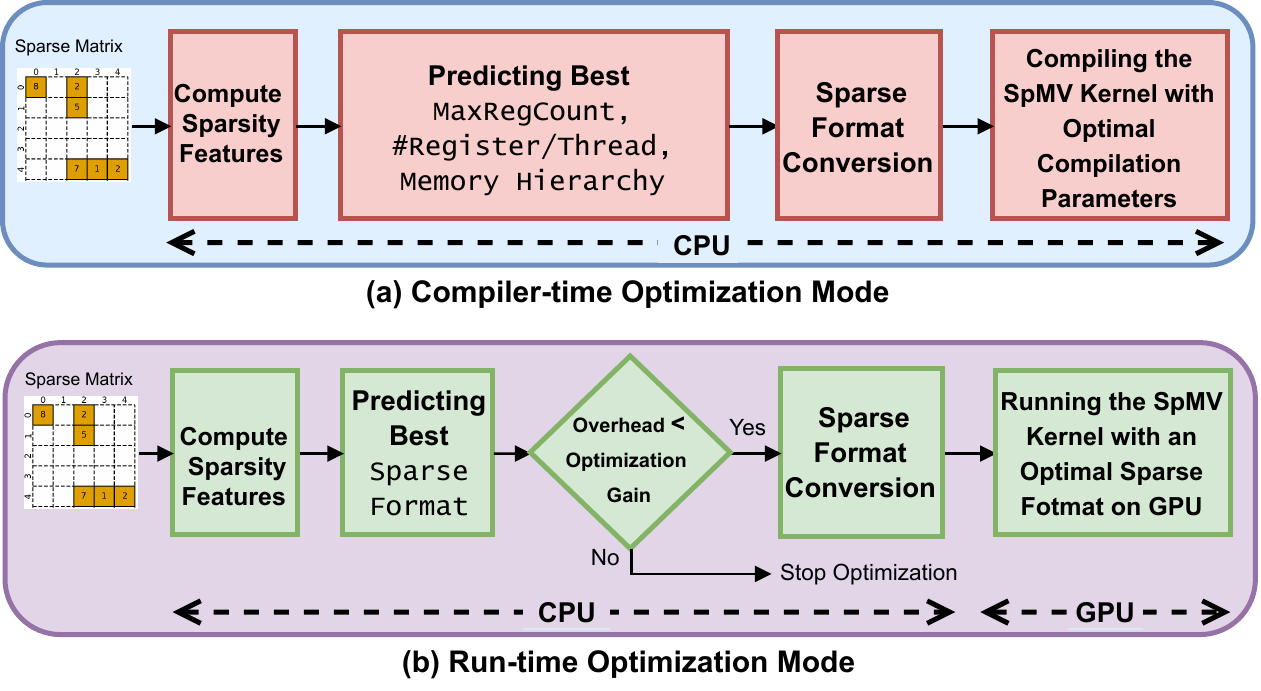}}
\caption{The overview of the proposed Auto-SpMV optimization framework. (a) \textit{compile-time} Optimization Mode. (b) \textit{run-time} Optimization Mode.}
\label{fig:method:Overview}
\end{center}
\end{figure}

As shown in Figure~\ref{fig:method:Overview}, Auto-SpMV optimizes SpMV kernels in the \textit{compile-time} and \textit{run-time} modes for various optimization objectives. 
In the \textit{compile-time} mode (Section~\ref{sec:Method:Offline}), the compiler configuration parameters are optimized.
In the \textit{run-time} mode (Section~\ref{sec:Method:Online}), Auto-SpMV optimizes the sparse format of the sparse input matrix.
Both optimization modes are conducted on the CPU.
In the rest of this section, we present the details of optimization modes, the sparsity features extracted from input matrices, and the Auto-SpMV performance prediction pipeline.

\subsection{\textit{Compile-time} Optimization Mode}
\label{sec:Method:Offline}

The aim of \textit{compile-time} optimization mode is to achieve the maximum GPU efficiency by tweaking the CUDA\textsuperscript{®} compilation parameters.
As shown in Figure~\ref{fig:method:Overview}, the \textit{compile-time} optimization mode includes the following steps:

\begin{enumerate}
    \item Compute the sparsity features.
    \item Predict the optimal GPU compilation parameters.
    \item Convert the sparse input matrix to CSR storage format (as default format) and compile CSR SpMV kernel with the optimal CUDA\textsuperscript{®} compilation parameters.
\end{enumerate}

We do not report the optimization overhead for the \textit{compile-time} optimization mode since the whole procedure is performed at the compilation time.

\subsection{\textit{Run-time} Optimization Mode}
\label{sec:Method:Online}

The aim of the \textit{run-time} optimization mode is to find the best-performing SpMV kernel and its sparse format for a sparse input matrix at run time.
Given a sparse matrix in a default sparse format with the default compilation parameters, the following steps are performed to predict its best sparse format:

\begin{enumerate}
    \item Compute the sparsity features.
    \item Predict the optimal sparse format for the given sparse matrix.
    \item Estimate the optimization overhead.
    \item Convert the sparse input matrix to the predicted best sparse format if the optimization benefit is higher than the estimated optimization overhead.
\end{enumerate}

With the aim of preventing unnecessary conversions for applications that are not composed of iterative solvers, Auto-SpMV leverages machine learning-based models for estimating the overhead of feature extraction and conversion to determine whether it is worthwhile to perform optimizations and tolerate the introduced overheads by subtracting estimated overhead (Section~\ref{sec:Results:Overhead}) from the predicted gain.
Figure~\ref{fig:overhead_prediction} shows the performance of estimating feature extraction and conversion overheads during the \textit{run-time} optimization mode.
Results show that Auto-SpMV accurately estimates the overhead of feature extraction and conversion at run time. 

In our work, we used the implementations of the SpMV kernel under CSR, ELL, BELL, and SELL formats.
The dataset used includes 30 sparse matrices from the SuiteSparse matrix collection \cite{davis2011university} (Section~\ref{sec:Setup:Benchmark})
Computing sparsity features, overhead/sparse format predictions, and conversion to the predicted best sparse format are done at run time of the application on the CPU. 
The overhead of online sparse format optimization is discussed in Section~\ref{sec:Results:Overhead}.

\begin{table}[htbp]
\resizebox{\columnwidth}{!}{
\begin{tabular}{cc}
    \begin{tikzpicture}
      \begin {axis}[
        width=\columnwidth,
        title style={at={(0.5,1.07)},anchor=north,yshift=-0.1},
        title =  \textbf{\huge Predicting Conversion Overhead},
        xmin=0, xmax=35,
        ymin=0, ymax=35,
        grid=major, 
        grid style={dashed,gray!40},
        xlabel= Actual (test data), 
        ylabel= Prediction (test data),
        legend style={at={(0.5,-0.2)},anchor=north,
        axis background/.style={fill=green!20},nodes={scale=1.3, transform shape}}, 
        legend pos=north west,
        label style={font=\huge},
                    tick label style={font=\Large} 
        ]
       \addplot [red, only marks, ultra  thick, mark options={scale=4}]
        table[x=Step_C,y=Value_C,col sep=comma] {RF_Acc._Predictor.csv}; 
        \addplot [thin, mark = none, blue, ultra thick ] coordinates {(0,0) (150,150)};
        \addlegendentry{\Large MSE=0.44, R$^{2}$=99.28\%}
\addlegendentry{\Large x=y}
      \end{axis}
    \end{tikzpicture}
&
    \begin{tikzpicture}
      \begin {axis}[
        width=\columnwidth,
        title style={at={(0.5,1.07)},anchor=north,yshift=-0.1},
        title =  \textbf{\huge Predicting Feature Extraction Overhead},
        xmin=0, xmax=25,
        ymin=0, ymax=25,
        grid=major, 
        grid style={dashed,gray!40},
        xlabel= Actual (test data), 
        ylabel= Prediction (test data),
        legend style={at={(0.5,-0.2)},anchor=north,
        axis background/.style={fill=green!20},nodes={scale=1.3, transform shape}}, 
        legend pos=north west,
        label style={font=\huge},
                    tick label style={font=\Large} 
        ]
       \addplot [red, only marks, ultra  thick, mark options={scale=4}]
        table[x=Step_F,y=Value_F,col sep=comma] {RF_Acc._Predictor.csv}; 
        \addplot [thin, mark = none, blue, ultra thick ] coordinates {(0,0) (160,160)};
        \addlegendentry{\Large MSE=0.36, R$^{2}$=99.18\%}
\addlegendentry{\Large x=y}
      \end{axis}
    \end{tikzpicture}
\end{tabular}}

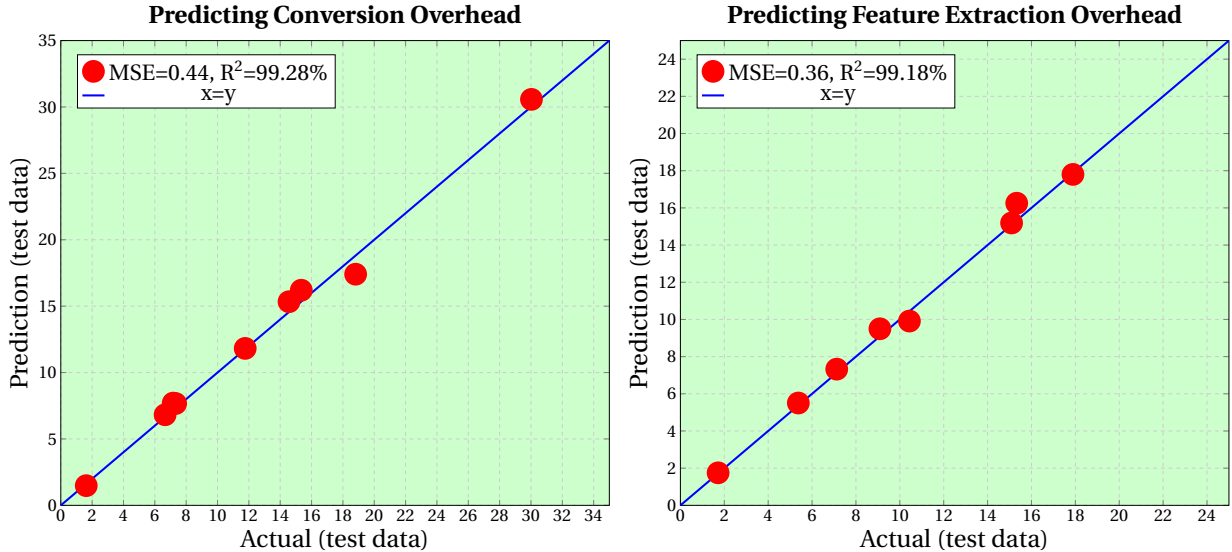
\captionof{figure}{The performance of estimating feature extraction and conversion overheads during the online optimization mode.} 
\label{fig:overhead_prediction}
\end{table}

\subsection{Predicting Optimal Configuration Parameters}
\label{sec:Method:ML}

In general, we have two alternatives to learn the best configuration parameters: 

\begin{enumerate}
    \item Multi-class classification: train a classifier that learns the optimal configuration parameters based on extracted sparsity features.
    Note that different classifiers should be trained for different optimization objectives.
    
    \item Regression: first, train different regression models for estimating different optimization objectives of the SpMV kernel.
    Then, select the optimal configuration parameters with the best estimations.
\end{enumerate}
Auto-SpMV relies on the multi-class classification approach. 
However, we also provide the results of regression models that estimate different optimization objectives of the SpMV kernel (Section~\ref{sec:Results:Prediction:Regression}). 
Using a machine learning algorithm, we can predict the best configuration parameters and sparse format of SpMV kernels by the following steps:

\begin{enumerate}
    \item \textbf{Step 1.} Extracting sparsity features of the sparse matrices that can represent the characteristics of running an SpMV kernel on GPU.
    
    \item \textbf{Step 2.} Constructing a learning dataset of diverse sparse matrices. 
    
    \item \textbf{Step 3.} Fine-tuning machine learning algorithms to provide the most accurate predictions.
\end{enumerate}

In this paper, we utilize supervised learning algorithms for both classification and regression tasks.
Further, we employed Optuna, a state-of-the-art hyperparameter optimization library for AutoML \cite{akiba2019optuna}.
Using Bayesian optimization \cite{shahriari2015taking} as the search method, Optuna automatically selects appropriate hyperparameters for a given dataset.

The main shortcomings of prior studies are 1) relying on one single learning model (e.g., BestSF only uses an SVM model \cite{[31]}), and 2) constructing a relatively small-scale optimization space (e.g., \cite{tsai2020sparse} only contains two optimization candidates). 
On the other hand, Auto-SpMV constructs a large-scale dataset containing four optimization objectives and fine-tunes six different learning models (nearest centroid, decision tree, non-linear SVM, gradient boosting, random forest, and multi-layer perceptron) using Optuna to provide the most accurate results.
Table~\ref{tab:learnig_fine-tuning} provides the hyperparameters and the corresponding range used to fine-tune the learning models. 
Section~\ref{sec:Setup} presents the specification of the learning models after fine-tuning. 

\begin{table*}[htbp]
\centering
\caption{The range of hyperparameters of each learning model selected for fine-tuning. }
\label{tab:learnig_fine-tuning}
\resizebox{\textwidth}{!}{%
\begin{tabular}{cc}
\hline
\textbf{Machine Learning Model} & \textbf{hyperparameters}   \\ \hline
\multirow{ 2}{*}{MLP} & hidden layer size:\{20, 50, 100, 150, 200\}, \#layer size:\{1, 2, 3, 4, 5, 10\},  \\ 
& Activation Function:\{"Identity", "Logistic", "Tanh", "ReLU"\}
\\ \hline

Decision Tree                                         & criterion:\{"gini", “entropy", "log\_loss"\}, splitter:\{"best", "random"\}   \\ \hline
Nearest Centroid Classifier                           & metric:\{"manhattan", "euclidean", "minkowski"\}             \\ \hline
Non-linear SVM                                        & kernel: \{"linear", "poly", "rbf", "sigmoid", "precomputed"\} \\ \hline
Gradient Boosting                                     & \#Estimators: {50, 100, 150, 200}, learning rate: {0.1, 0.01, 0.001}                                                          \\ \hline
Random Forest                                         & criterion: \{"gini", "entropy", "log\_loss"\}    \\  \hline         
\end{tabular}%
}
\end{table*}

\subsection{Sparsity Features}
\label{sec:Method:Sparsity_Features}

In order to determine the optimal configuration parameters in both optimization modes, Auto-SpMV first needs to extract eight different sparsity features from the sparse input matrix.
These sparsity features are briefly described in Table~\ref{tab:Matrix_Features}.

\begin{table}[htbp]
\centering
\caption{Sparsity Features Selected by Auto-SpMV.}
\label{tab:Matrix_Features}
\resizebox{\columnwidth}{!}{%
\begin{tabular}{cc}
\hline
\textbf{Feature} & \multicolumn{1}{c}{\textbf{Definition}}               \\\hline
\texttt{n} & The number of rows \\
\texttt{nnz} & The number of non-zero elements  \\
\texttt{Avg\_nnz} & The average number of non-zero elements in rows  \\
\texttt{Var\_nnz} & The variance of non-zero elements in rows \\
\texttt{ELL\_ratio} & The ratio of non-zero elements to the size of the matrix in the ELL format \\
\texttt{Median} & The median of non-zero elements in rows   \\
\texttt{Mode}  & The mode of non-zero elements in rows                     \\
\texttt{Std\_nnz} & The standard deviation of non-zero   elements in rows\\ \hline
\end{tabular}
}
\end{table}

We selected features based on 1) having the minimum computation overhead at run-time and 2) the performance impact reported by related papers \cite{[31]}.
The complexity of GPUs and the wide range of sparsity patterns make it impossible to explain exactly how each sparsity feature impacts the performance of the SpMV kernel on GPUs.
Nevertheless, we list the following general observations regarding the performance of different sparse formats:
\begin{enumerate}
    \item The number of rows (\texttt{n}) and the number of non-zero elements (\textbf{nnz}) characterize the volume of the SpMV kernel computation running on GPU.
    
    \item The performance of the SpMV on GPU, depending on the used sparse format, is sensitive to the dispersion level of the non-zero elements on the rows of the sparse matrix. To characterize this sensitivity, we included \texttt{Avg\_nz}, \texttt{Var\_nnz}, \texttt{Std\_nnz} sparsity features.
    
    \item CSR SpMV kernel gives good performance for matrices with regular distribution of the non-zero elements, characterized by large values of \texttt{Avg\_nz} and relatively small values of \texttt{Std\_nnz}, for which these kernels do not suffer from unbalanced distribution among the threads.
    
    \item Small values of \texttt{ELL\_ratio} is a good indicator for the existence of many blocks of non-zero elements in the matrix which favors the ELL format because, in this case, ELL will less suffer from the problem of excessive zero padding that introduces more computation than needed.
\end{enumerate}

\section{Experimental Setup}
\label{sec:Setup}

\subsection{Benchmark}
\label{sec:Setup:Benchmark}

SuiteSparse \cite{davis2011university} provides an extensive collection of sparse matrices.
SuiteSparse is derived from real-world applications, such as numerical linear algebra, computational fluid dynamics, computer graphics, and financial modeling.
We select 30 sparse matrices from the SuiteSparse library on the basis of three criteria: 1) matrices with a wide range of dimension size (14,340$<$\texttt{n}$<$1,489,752), 2) matrices with a wide range of non-zero elements (800,800$<$\texttt(nnz)$<$19,235,140), and 3) matrices with the minimum similarity between their sparsity features to cover more diverse benchmarks.
Figure~\ref{fig:dataset_distribution} presents the distribution of sparsity features for the selected matrices sorted by the ascending order of \emph{nnz}.
Figure~\ref{fig:correlation} shows the Pearson correlation of sparsity features.
The results show that there exists a low degree of correlation among the sparsity features of selected matrices.

\begin{table*}[htbp]
\centering
\resizebox{\linewidth}{!}{
\begin{tabular}{cc}

\begin{tikzpicture}

\begin{axis}[
    title={\Huge \textbf{\texttt{n}}},
    width=\columnwidth,
    height=0.4\columnwidth,
    font=\large ,
    ylabel={\Large {Value}},
    xlabel={\Large {Benchmark}},
    symbolic x coords={$bcsstk32$,$pkustk04$,$apache2$,$x104$,$crankseg\_1$,$pwtk$,$crankseg\_2$,$Si87H76$,$Chevron3$,$cant$,$consph$,$Chevron4$,$rim$,$viscorocks$,$amazon0601$,$xenon2$,$il2010$,$wiki\rnumber talk$,$af\_shell6$,$test1$,$rgg\_n\_2\_20\_s0$,$shar\_te2\rnumber b3$,$kim2$,$human\_gene2$,$Hardesty1$,$Hamrle3$,$CurlCurl\_3$,$eu\rnumber 2005$,$atmosmodm$,$parabolic\_fem$},
    xtick=data,
    xticklabel style={rotate=90},
    nodes near coords align={vertical},
    ]
\addplot [scatter, only marks, mark size=4pt] coordinates {($bcsstk32$,44609) ($pkustk04$,55590) ($apache2$,715176) ($x104$,108384) ($crankseg\_1$,52804) ($pwtk$,217918) ($crankseg\_2$,63838) ($Si87H76$,240369) ($Chevron3$,381381) ($cant$,62451) ($consph$,83334)($Chevron4$,711450) ($rim$,22560) ($viscorocks$,37762) ($amazon0601$,403394) ($xenon2$,157464) ($il2010$,451554) ($wiki\rnumber talk$,1140149) ($af\_shell6$,504855) ($test1$,392908) ($rgg\_n\_2\_20\_s0$,1048576) ($shar\_te2\rnumber b3$,200200) ($kim2$,456976) ($human\_gene2$,14340) ($Hardesty1$,938905) ($Hamrle3$,1447360) ($CurlCurl\_3$,1219574) ($eu\rnumber 2005$,862664) ($atmosmodm$,1489752)($parabolic\_fem$,525825) };

\end{axis}
\end{tikzpicture}

&

\begin{tikzpicture}
\begin{axis}[
    title={\Huge \textbf{\texttt{nnz}}},
    width=\columnwidth,
    height=0.4\columnwidth,
    font=\large,
    ylabel={\Large {Value}},
    xlabel={\Large {Benchmark}},
    symbolic x coords={$bcsstk32$,$pkustk04$,$apache2$,$x104$,$crankseg\_1$,$pwtk$,$crankseg\_2$,$Si87H76$,$Chevron3$,$cant$,$consph$,$Chevron4$,$rim$,$viscorocks$,$amazon0601$,$xenon2$,$il2010$,$wiki\rnumber talk$,$af\_shell6$,$test1$,$rgg\_n\_2\_20\_s0$,$shar\_te2\rnumber b3$,$kim2$,$human\_gene2$,$Hardesty1$,$Hamrle3$,$CurlCurl\_3$,$eu\rnumber 2005$,$atmosmodm$,$parabolic\_fem$},
    xtick=data,
    xticklabel style={rotate=90},
    nodes near coords align={vertical},
    ]
\addplot [scatter, only marks, mark size=4pt] coordinates {($bcsstk32$, 1029655) ($pkustk04$, 2137125) ($apache2$, 2766523) ($x104$, 5138004) ($crankseg\_1$, 5333507) ($pwtk$, 5926171) ($crankseg\_2$, 7106348) ($Si87H76$, 5451000) ($Chevron3$, 3413113) ($cant$, 2034917) ($consph$, 3046907) ($Chevron4$, 6376412) ($rim$, 1014951) ($viscorocks$, 1162244) ($amazon0601$, 3387388) ($xenon2$, 3866688) ($il2010$, 1082232) ($wiki\rnumber talk$, 3309592) ($af\_shell6$, 9046865) ($test1$, 12968200) ($rgg\_n\_2\_20\_s0$, 6891620) ($shar\_te2\rnumber b3$, 800800) ($kim2$, 11330020) ($human\_gene2$, 9041364) ($Hardesty1$, 6539157) ($Hamrle3$, 5514242) ($CurlCurl\_3$,7382096 ) ($eu\rnumber 2005$, 19235140) ($atmosmodm$, 10319760)($parabolic\_fem$, 2100225) };

\end{axis}
\end{tikzpicture}

\\

\begin{tikzpicture}
\begin{axis}[
    title={\Huge \textbf{\texttt{avr\_nnz}}},
    width=\columnwidth,
    height=0.4\columnwidth,
    font=\large,
    ymode=log,
    ylabel={\Large {Value (Log Scale)}},
    xlabel={\Large {Benchmark}},
    symbolic x coords={$bcsstk32$,$pkustk04$,$apache2$,$x104$,$crankseg\_1$,$pwtk$,$crankseg\_2$,$Si87H76$,$Chevron3$,$cant$,$consph$,$Chevron4$,$rim$,$viscorocks$,$amazon0601$,$xenon2$,$il2010$,$wiki\rnumber talk$,$af\_shell6$,$test1$,$rgg\_n\_2\_20\_s0$,$shar\_te2\rnumber b3$,$kim2$,$human\_gene2$,$Hardesty1$,$Hamrle3$,$CurlCurl\_3$,$eu\rnumber 2005$,$atmosmodm$,$parabolic\_fem$},
    xtick=data,
    xticklabel style={rotate=90},
    nodes near coords align={vertical},
    ]
\addplot [scatter, only marks, mark size=4pt] coordinates {($bcsstk32$, 0) ($pkustk04$, 38.444416) ($apache2$, 3.868311) ($x104$, 47.40556) ($crankseg\_1$, 101.005737) ($pwtk$, 27.1945) ($crankseg\_2$, 	111.318459) ($Si87H76$, 22.677633) ($Chevron3$, 8.949352) ($cant$, 32.584217) ($consph$, 36.562592) ($Chevron4$, 8.962558) ($rim$, 44.988964) ($viscorocks$, 30.778137) ($amazon0601$,8.39722) ($xenon2$,24.556013 ) ($il2010$,2.396683 ) ($wiki\rnumber talk$, 2.902771) ($af\_shell6$,17.919729) ($test1$,33.005692) ($rgg\_n\_2\_20\_s0$,6.572361) ($shar\_te2\rnumber b3$, 4) ($kim2$, 24.793468) ($human\_gene2$, 630.499573) ($Hardesty1$, 6.964663) ($Hamrle3$, 3.809862) ($CurlCurl\_3$,6.053012 ) ($eu\rnumber 2005$,22.289568) ($atmosmodm$,6.927166)($parabolic\_fem$,3.994152) };

\end{axis}
\end{tikzpicture}

&

\begin{tikzpicture}
\begin{axis}[
    title={\Huge \textbf{\texttt{var\_nnz}}},
    width=\columnwidth,
    height=0.4\columnwidth,
    font=\large,
    ymode=log,
    ylabel={\Large {Value (Log Scale)}},
    xlabel={\Large {Benchmark}},
    symbolic x coords={$bcsstk32$,$pkustk04$,$apache2$,$x104$,$crankseg\_1$,$pwtk$,$crankseg\_2$,$Si87H76$,$Chevron3$,$cant$,$consph$,$Chevron4$,$rim$,$viscorocks$,$amazon0601$,$xenon2$,$il2010$,$wiki\rnumber talk$,$af\_shell6$,$test1$,$rgg\_n\_2\_20\_s0$,$shar\_te2\rnumber b3$,$kim2$,$human\_gene2$,$Hardesty1$,$Hamrle3$,$CurlCurl\_3$,$eu\rnumber 2005$,$atmosmodm$,$parabolic\_fem$},
    xtick=data,
    xticklabel style={rotate=90},
    nodes near coords align={vertical},
    ]
\addplot [scatter, only marks, mark size=4pt] coordinates {($bcsstk32$, 102.107666) ($pkustk04$, 11353.49902) ($apache2$, 0.115275) ($x104$,312.976746) ($crankseg\_1$,2141.539063) ($pwtk$,31.569984) ($crankseg\_2$,2339.70459) ($Si87H76$,532.053589) ($Chevron3$,0.350953)  ($cant$, 53.81852) ($consph$, 118.777771) ($Chevron4$, 0.259386) ($rim$, 706.229187) ($viscorocks$, 59.846333) ($amazon0601$, 7.785032) ($xenon2$, 16.56608) ($il2010$, 4.362809) ($wiki\rnumber talk$, 28431.37109) ($af\_shell6$, 27.694305) ($test1$, 102.978394) ($rgg\_n\_2\_20\_s0$, 6.577159) ($shar\_te2\rnumber b3$, 0) ($kim2$,3.599854 ) ($human\_gene2$, 731737.875) ($Hardesty1$, 0.177319) ($Hamrle3$, 2.396748) ($CurlCurl\_3$, 5.731723) ($eu\rnumber 2005$,860.458008) ($atmosmodm$, 0.068635)($parabolic\_fem$, 4.230037) };

\end{axis}
\end{tikzpicture}

\\

\begin{tikzpicture}
\begin{axis}[
    title={\Huge \textbf{\texttt{ell\_ratio}}},
    width=\columnwidth,
    height=0.4\columnwidth,
    font=\large,
    ylabel={\Large {Value}},
    xlabel={\Large {Benchmark}},
    symbolic x coords={$bcsstk32$,$pkustk04$,$apache2$,$x104$,$crankseg\_1$,$pwtk$,$crankseg\_2$,$Si87H76$,$Chevron3$,$cant$,$consph$,$Chevron4$,$rim$,$viscorocks$,$amazon0601$,$xenon2$,$il2010$,$wiki\rnumber talk$,$af\_shell6$,$test1$,$rgg\_n\_2\_20\_s0$,$shar\_te2\rnumber b3$,$kim2$,$human\_gene2$,$Hardesty1$,$Hamrle3$,$CurlCurl\_3$,$eu\rnumber 2005$,$atmosmodm$,$parabolic\_fem$},
    xtick=data,
    xticklabel style={rotate=90},
    nodes near coords align={vertical},
    ]
\addplot [scatter, only marks, mark size=4pt] coordinates {($bcsstk32$, 0.163701) ($pkustk04$, 0.009153) ($apache2$, 0.967078) ($x104$, 0.23238) ($crankseg\_1$, 0.385518) ($pwtk$, 0.151081) ($crankseg\_2$, 0.37481) ($Si87H76$, 0.096913) ($Chevron3$, 0.994372) ($cant$, 0.814605) ($consph$,0.553979 ) ($Chevron4$, 	0.99584) ($rim$, 0.401687) ($viscorocks$, 0.732813) ($amazon0601$, 0.839722) ($xenon2$, 0.909482) ($il2010$, 0.029959) ($wiki\rnumber talk$, -0.002299) ($af\_shell6$, 0.511992) ($test1$, 0.3549) ($rgg\_n\_2\_20\_s0$, 0.285755) ($shar\_te2\rnumber b3$, 1) ($kim2$,0.991739 ) ($human\_gene2$, 0.113114) ($Hardesty1$, 0.994952) ($Hamrle3$, 0.634977) ($CurlCurl\_3$, 0.550274) ($eu\rnumber 2005$, 0.011114) ($atmosmodm$,0.989595 )($parabolic\_fem$, 0.570593) };

\end{axis}
\end{tikzpicture}

&

\begin{tikzpicture}
\begin{axis}[
    title={\Huge \textbf{\texttt{median}}},
    width=\columnwidth,
    height=0.4\columnwidth,
    font=\large,
    ymode=log,
    ylabel={\Large {Value (Log Scale)}},
    xlabel={\Large {Benchmark}},
    symbolic x coords={$bcsstk32$,$pkustk04$,$apache2$,$x104$,$crankseg\_1$,$pwtk$,$crankseg\_2$,$Si87H76$,$Chevron3$,$cant$,$consph$,$Chevron4$,$rim$,$viscorocks$,$amazon0601$,$xenon2$,$il2010$,$wiki\rnumber talk$,$af\_shell6$,$test1$,$rgg\_n\_2\_20\_s0$,$shar\_te2\rnumber b3$,$kim2$,$human\_gene2$,$Hardesty1$,$Hamrle3$,$CurlCurl\_3$,$eu\rnumber 2005$,$atmosmodm$,$parabolic\_fem$},
    xtick=data,
    xticklabel style={rotate=90},
    nodes near coords align={vertical},
    ]
\addplot [scatter, only marks, mark size=4pt] coordinates {($bcsstk32$, 24) ($pkustk04$, 18) ($apache2$, 4) ($x104$, 41) ($crankseg\_1$, 99) ($pwtk$, 27) ($crankseg\_2$, 109) ($Si87H76$, 19) ($Chevron3$, 9) ($cant$, 37) ($consph$, 40) ($Chevron4$, 9) ($rim$, 36) ($viscorocks$, 26) ($amazon0601$, 10) ($xenon2$, 27) ($il2010$, 2) ($wiki\rnumber talk$, 0) ($af\_shell6$, 15) ($test1$, 33) ($rgg\_n\_2\_20\_s0$,6 ) ($shar\_te2\rnumber b3$,4 ) ($kim2$, 25) ($human\_gene2$, 282) ($Hardesty1$, 7) ($Hamrle3$, 5) ($CurlCurl\_3$, 6) ($eu\rnumber 2005$,16 ) ($atmosmodm$, 7)($parabolic\_fem$, 4) };

\end{axis}
\end{tikzpicture}

\\

\begin{tikzpicture}
\begin{axis}[
    title={\Huge \textbf{\texttt{mode\_element}}},
    width=\columnwidth,
    height=0.4\columnwidth,
    font=\large,
    ylabel={\Large {Value}},
    xlabel={\Large {Benchmark}},
    symbolic x coords={$bcsstk32$,$pkustk04$,$apache2$,$x104$,$crankseg\_1$,$pwtk$,$crankseg\_2$,$Si87H76$,$Chevron3$,$cant$,$consph$,$Chevron4$,$rim$,$viscorocks$,$amazon0601$,$xenon2$,$il2010$,$wiki\rnumber talk$,$af\_shell6$,$test1$,$rgg\_n\_2\_20\_s0$,$shar\_te2\rnumber b3$,$kim2$,$human\_gene2$,$Hardesty1$,$Hamrle3$,$CurlCurl\_3$,$eu\rnumber 2005$,$atmosmodm$,$parabolic\_fem$},
    xtick=data,
    xticklabel style={rotate=90},
    nodes near coords align={vertical},
    ]
\addplot [scatter, only marks, mark size=4pt] coordinates {($bcsstk32$, 25) ($pkustk04$, 1) ($apache2$, 3) ($x104$,37 ) ($crankseg\_1$,85 ) ($pwtk$,25 ) ($crankseg\_2$, 97) ($Si87H76$, 19) ($Chevron3$, 2) ($cant$, 37) ($consph$, 41)($Chevron4$, 2) ($rim$, 30) ($viscorocks$, 26) ($amazon0601$,3 ) ($xenon2$, 22) ($il2010$, 1) ($wiki\rnumber talk$, 1) ($af\_shell6$, 11) ($test1$, 33) ($rgg\_n\_2\_20\_s0$, 6) ($shar\_te2\rnumber b3$, 1) ($kim2$, 6) ($human\_gene2$, 1) ($Hardesty1$, 1) ($Hamrle3$, 5) ($CurlCurl\_3$, 7) ($eu\rnumber 2005$,1 ) ($atmosmodm$, 6)($parabolic\_fem$, 5) };

\end{axis}
\end{tikzpicture}

&

\begin{tikzpicture}
\begin{axis}[
    title={\Huge \textbf{\texttt{SD\_nnz}}},
    width=\columnwidth,
    height=0.4\columnwidth,
    font=\large,
    ymode=log,
    ylabel={\Large {Value (Log Scale)}},
    xlabel={\Large {Benchmark}},
    symbolic x coords={$bcsstk32$,$pkustk04$,$apache2$,$x104$,$crankseg\_1$,$pwtk$,$crankseg\_2$,$Si87H76$,$Chevron3$,$cant$,$consph$,$Chevron4$,$rim$,$viscorocks$,$amazon0601$,$xenon2$,$il2010$,$wiki\rnumber talk$,$af\_shell6$,$test1$,$rgg\_n\_2\_20\_s0$,$shar\_te2\rnumber b3$,$kim2$,$human\_gene2$,$Hardesty1$,$Hamrle3$,$CurlCurl\_3$,$eu\rnumber 2005$,$atmosmodm$,$parabolic\_fem$},
    xtick=data,
    xticklabel style={rotate=90},
    nodes near coords align={vertical},
    ]
\addplot [scatter, only marks, mark size=4pt] coordinates {($bcsstk32$, 10.104834) ($pkustk04$, 106.552803) ($apache2$, 0.339522) ($x104$, 17.691149) ($crankseg\_1$, 46.276768) ($pwtk$, 5.618717) ($crankseg\_2$, 48.370495) ($Si87H76$, 23.066286) ($Chevron3$, 0.592413) ($cant$, 7.336111) ($consph$, 10.898521) ($Chevron4$, 0.5093) ($rim$, 26.574972) ($viscorocks$, 7.736041) ($amazon0601$, 2.790167) ($xenon2$, 4.070145) ($il2010$, 2.088734) ($wiki\rnumber talk$, 168.616043) ($af\_shell6$, 5.262538) ($test1$, 10.147827) ($rgg\_n\_2\_20\_s0$, 2.5645970) ($shar\_te2\rnumber b3$, 0) ($kim2$, 1.897328) ($human\_gene2$,855.416809) ($Hardesty1$, 0.421092) ($Hamrle3$, 1.548143) ($CurlCurl\_3$, 2.394102) ($eu\rnumber 2005$,29.333565) ($atmosmodm$,0.261983 )($parabolic\_fem$,2.056705) };

\end{axis}
\end{tikzpicture}

\end{tabular}
}

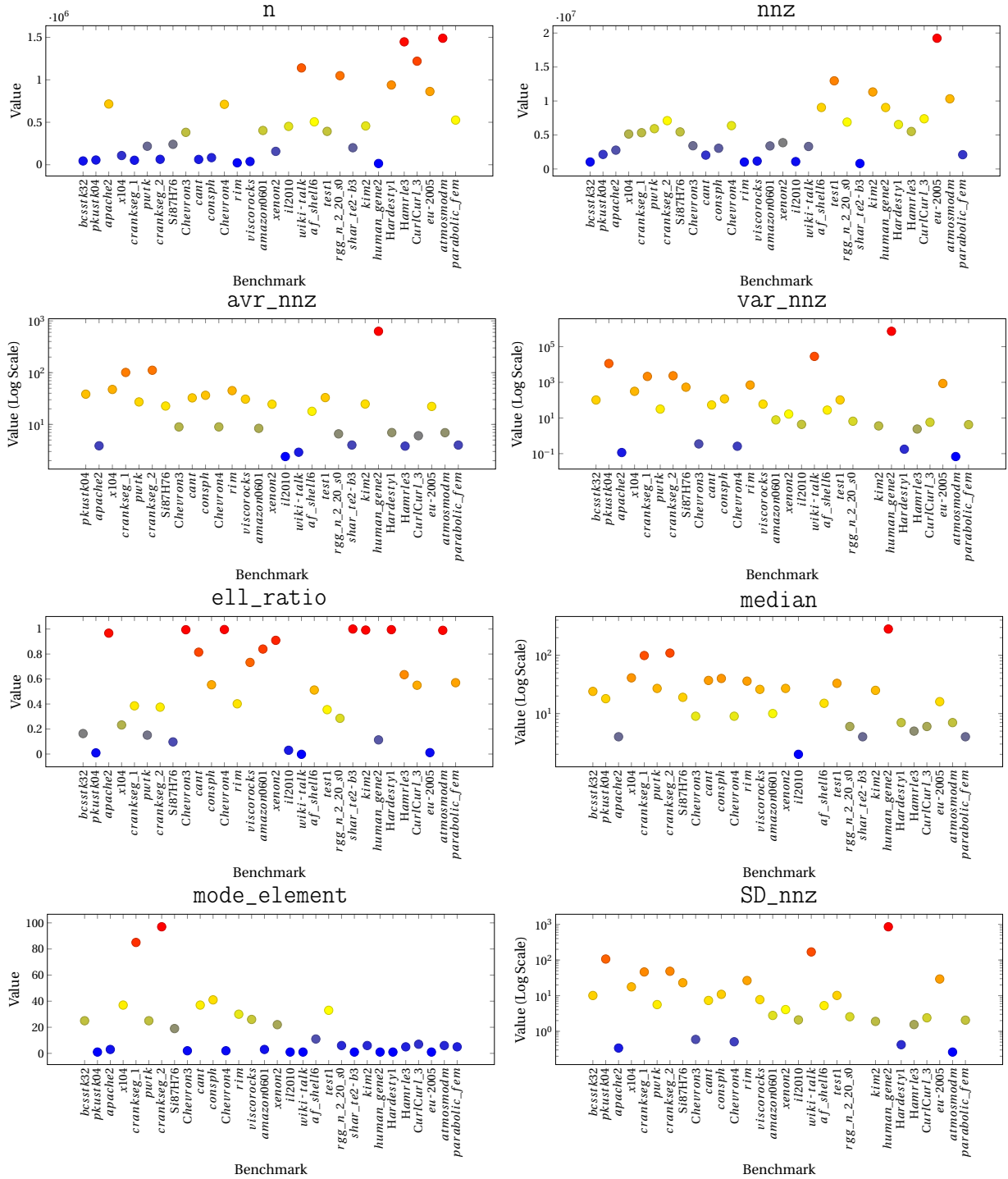
\captionof{figure}{The distribution of the structural features over the studied matrices sorted by increasing values of \emph{nnz}. We see that the selected matrices cover a wide range of sparsity features.}
\label{fig:dataset_distribution}
\end{table*}

\begin{table}[hbpt]
\centering
\resizebox{0.75\columnwidth}{!}{
\begin{tabular}{c}

\begin{tikzpicture}
\begin{customlegend}[legend columns=4,legend style={text opacity = 1,row sep=0pt, font=\fontsize{12}{8}\selectfont, column sep=2ex},
        legend entries={\textbf{\texttt{n}},\textbf{\texttt{nnz}},\textbf{\texttt{Avg\_nnz}},\textbf{\texttt{Var\_nnz}}, \textbf{\texttt{ELL\_ratio}},\textbf{\texttt{Median}}, \textbf{\texttt{Mode}}, \textbf{\texttt{SD\_nnz}}}]
        
        \addlegendimage{mark=square*, mark size=7pt, only marks, blue, thick, draw=black}
        \addlegendimage{mark=square*, mark size=7pt, only marks, red, thick, draw=black}
        \addlegendimage{mark=square*, mark size=7pt, only marks, green, thick, draw=black}
        \addlegendimage{mark=square*, mark size=7pt, only marks, cyan, thick, draw=black}
        \addlegendimage{mark=square*, mark size=7pt, only marks, pink, thick, draw=black}
         \addlegendimage{mark=square*, mark size=7pt, only marks, violet, thick, draw=black}
        \addlegendimage{mark=square*, mark size=7pt, only marks, , teal, draw=black}
        \addlegendimage{mark=square*, mark size=7pt, only marks, orange, thick, draw=black}
        \end{customlegend}
\end{tikzpicture}

\\
\begin{tikzpicture}
\tkzKiviatDiagram[scale=0.75,label distance=1cm,
        gap     = 1,  
        lattice = 4]{\textbf{\texttt{n}}, \textbf{\texttt{nnz}}, \textbf{\texttt{Avg\_nnz}}, \textbf{\texttt{Var\_nnz}}, \textbf{\texttt{ELL\_ratio}}, \textbf{\texttt{Median}}, \textbf{\texttt{Mode}}, \textbf{\texttt{SD\_nnz}}}

\tkzKiviatLine[ultra thick,color=blue,mark=none](4,	2.367116706, 1.2, 1.32, 2.182782668, 1.08, 1.11, 1.32)
\tkzKiviatLine[ultra thick,color=red,mark=none](2.367116706, 4, 2.176493498,	2.155832719, 1.39, 2.168477063,1.46, 2.138180339)
\tkzKiviatLine[ultra thick,color=green,mark=none](1.2, 2.18, 4, 3.97, 1.25, 3.96, 2.08, 3.96)
\tkzKiviatLine[ultra thick,color=cyan, dashed,mark=none](1.32, 2.155832719, 3.971912615, 4, 1.27, 3.879877602, 1.36, 3.981809921)
\tkzKiviatLine[ultra thick,color=pink,mark=none](2.182782668, 1.38,1.24, 1.27, 4, 1.23, 1.35, 1.17)
\tkzKiviatLine[ultra thick,color=violet,mark=none](1.08, 2.168477063, 3.963883327, 3.879877602, 1.23, 4, 2.335632022, 3.874765163)
\tkzKiviatLine[ultra thick,color=teal,mark=none](1.01, 1.47, 2.079632872, 1.16, 1.35, 2.33, 4, 1.39) 
\tkzKiviatLine[ultra thick,color=orange,mark=none](1.31, 2.138180339, 3.960000525, 3.981809921, 1.17, 3.874765163, 1.39, 4) 
\tkzKiviatGrad[unity=50](0)
\end{tikzpicture}

\end{tabular}
}
\captionof{figure}{The Pearson correlation (\%) of sparsity features of selected benchmarks. We see that there is no correlation between them.}
\label{fig:correlation}
\end{table}

\subsection{Hardware Specification}
\label{sec:Setup:HW_Configuration}

In this paper, we utilize two different NVIDIA\textsuperscript{®} GPU architectures, Pascal and Turing, to evaluate the sensitivity of learning results to the variation of hardware devices.
Table~\ref{tab:hardware_spec} presents the specification of utilized hardware devices.
We use NVIDIA\textsuperscript{®} CUDA\textsuperscript{®} v. 11 to compile SpMV kernels on GPU. 

\begin{table}[htbp]
  \centering
  \caption{Hardware Specification.}
  \label{tab:hardware_spec}
  \resizebox{0.65\textwidth}{!}{
    \begin{tabular}{ccc}
    \hline
\textbf{GPU Device} & \textbf{Specification} & \textbf{Value} \\\hline 
   \multirow{3}{*}{GTX 1650-mobile} & Clock Frequency & 1.6 GHz  \\
   & On-chip Memory	& 4GB GDDR5X  \\ 
   & \# CUDA\textsuperscript{®} Core	& 896  \\ \cline{2-3}
    \multirow{3}{*}{GTX 1080} & Base Clock Frequency & 1.6 GHz  \\
   & On-chip Memory	& 8GB GDDR5X  \\
   & \# CUDA\textsuperscript{®} Core	& 2560  \\ \hline
   \textbf{CPU Device} & \textbf{Specification} & \textbf{Value} \\ \hline  \multirow{2}{*}{AMD\textsuperscript{®} Ryzen 5} & Clock Frequency & 3.6 GHz \\
   & Memory & 16 GB
   \\\hline
    \end{tabular}
    }
\end{table}

\subsection{GPU Performance Monitoring}
\label{sec:Setup:GPU_Monitoring}

In this paper, we consider four optimization objectives, including 1) latency, 2) energy consumption, 3) average power consumption, and 4) energy efficiency (MFLOPS/Watt).
The execution time and energy consumption are measured using the multi-threading technique by a separate thread running on the host CPU besides another thread which is responsible for allocating memory and launching the kernel on GPU.
For computing, the execution time, function \texttt{QueryPerformanceCounter()} of the NVIDIA\textsuperscript{®} Management Library (NVML) is utilized.
This function retrieves the current value of the performance counter, with a high-resolution timestamp ($<1us$).
Depending on the kernel complexity, We run each kernel 500-200000 times on GPU due to the required time intervals for GPU on-chip sensors in returning the measured instant power consumption \cite{burtscher2014measuring}.
The average of these runs is then reported as the kernel's execution time. 

We employ GPU power sensors to obtain energy and power consumption.
The thread running on the CPU queries the internal sensors via the NVML interface, which returns the power readings in milliwatts.
Energy and average power consumption are computed without taking into account idle power samples.
The arithmetic mean of power samples is considered as the average power consumption, and the integral of power samples over time samples is regarded as the energy consumption of an SpMV kernel.
Finally, energy efficiency (MFLOPS/Watt) is calculated by dividing the achieved mega floating-point operations per second (MFLOPS) by the average power consumption.

\subsection{Learning Algorithms}
\label{sec:Setup:Learning}

Table~\ref{tab:ML_Spec} presents the specification of the best learning models after fine-tuning.
The rest of the specification of learning models is set to the default hyperparameters of Scikit-learn machine learning library \cite{Scikit}.
We used 80\% of the dataset for training and 20\% for validating the prediction results.

\begin{table}[htbp]
\centering
\caption{Summarizing the specification of learning models after fine-tuning utilized for predicting the best hyperparameters (classification) and estimating various objective functions (regression) of SpMV kernels on GPU.}
\label{tab:ML_Spec}
\resizebox{0.9\columnwidth}{!}{
\begin{tabular}{ccc}
\hline
\multirow{10}{*}{\rotatebox{90}{\textbf{Classification}}} & \textbf{Learning Algorithm} & \textbf{hyperparameter \& Values}\\ \hline

& Nearest Centroid  & metric=manhattan, \#neighbors=5\\ \cline{2-3}

& Decision Tree  &  Depth = 13\\ \cline{2-3}

& Non-linear SVM & kernel=rbf, C=1.0, degree=3, gamma=scale\\ \cline{2-3} 

& Gradient boosting & loss=hinge, alpha=0.0001, \#epochs=1000\\ \cline{2-3}
 
& \multirow{2}{*}{Random Forest} &  \#estimators=100\\

& & The maximum Depth=15 \\ \cline{2-3}

& \multirow{3}{*}{Multi-layer Perceptron}  & \#layers=5, \#nodes/layer=100  \\
& & activation=ReLU, \#epochs= 200 \\
& & Optimizer=Adam and learning rate=0.001 \\\hline

\multirow{8}{*}{\rotatebox{90}{\textbf{Regression}}}& Bayesian Ridge & \#iter=300,  tol=0.001\\ \cline{2-3}
& Lasso & alpha=1.0 , \#epochs=1000\\ \cline{2-3}
& Random Forest& \#estimators=100, Depth=None \\
& Decision Tree & Depth=None \\ \cline{2-3}
& Lars & \#non-zero coefs=500, eps=2.22045e-16,\\\cline{2-3}
& \multirow{3}{*}{Multi-layer Perceptron}  & \#layers=5, \#nodes/layer=200  \\
& & activation=ReLU, \#epochs= 200 \\
& & Optimizer=Adam and learning rate=0.0001
\\ \hline
\end{tabular}}
\end{table}

\section{Experimental Results}
\label{sec:Results}

\subsection{Results of the \textit{Compile-time} Optimization Mode}
\label{sec:Results:Offline}

Figure~\ref{fig:result:offline} compares the performance improvement of Auto-SpMV in the \textit{compile-time} optimization mode over the default parameters with the CSR sparsity format for different optimization objectives.
The upper and lower bounds of each bar show the best and the worse results when tweaking  the thread block size as the programmer is responsible for selecting this parameter.   
Results show that Auto-SpMV achieves up to 51.9\%, 52\%, 33.2\%, and 53\% improvement for all selected sparse matrices if we optimize the compiler parameters in terms of latency, energy consumption, average power, and energy efficiency, respectively. 

\begin{table*}[htbp]
\centering
\resizebox{\textwidth}{!}{
\begin{tabular}{c}

\begin{tikzpicture}
\begin{axis}[ybar,
        title=Latency,
        width=\textwidth,
        height=0.25\columnwidth,
        ymin=0,
        ymax=60,
        bar width=6pt,
        enlargelimits=0.02,
        ylabel={Improvement (\%)},
        xticklabels=empty,
        minor x tick num = 1,
        minor tick length=0ex
        ]
\addplot[draw=black, fill=red,error bars/.cd,y dir=both,y explicit] 
    table[x index=0,y index=3,y error plus index=2,y error minus index=1, col sep=comma] {results_perf.csv};
    
\end{axis}
\end{tikzpicture}

\\
\begin{tikzpicture}
\begin{axis}[ybar,
        title=Energy Consumption,
        width=\textwidth,
        height=0.25\columnwidth,
        ymin=0,
        ymax=60,
        bar width=6pt,
        enlargelimits=0.02,
        ylabel={Improvement (\%)},
        xticklabels=empty,
        minor x tick num = 1,
        minor tick length=0ex
        ]
\addplot[draw=black, fill=teal,error bars/.cd,y dir=both,y explicit] 
    table[x index=0,y index=3,y error plus index=2,y error minus index=1, col sep=comma] {results_energy.csv};  
    
\end{axis}
\end{tikzpicture}
\\
\begin{tikzpicture}
\begin{axis}[ybar,
        title=Average Power Consumption,
        width=\textwidth,
        height=0.25\columnwidth,
        ymin=0,
        ymax=40,
        bar width=6pt,
        enlargelimits=0.02,
        ylabel={Improvement (\%)},
        xticklabels=empty,
        minor x tick num = 1,
        minor tick length=0ex
        ]
\addplot[draw=black, fill=orange,error bars/.cd,y dir=both,y explicit] 
    table[x index=0,y index=3,y error plus index=2,y error minus index=1, col sep=comma] {results_avg_power.csv}; 
    
\end{axis}
\end{tikzpicture}
\\
\begin{tikzpicture}
\begin{axis}[ybar,
        title=Energy Efficiency,
        width=\textwidth,
        height=0.25\columnwidth,
        ymin=0,
        ymax=60,
        bar width=6pt,
        enlargelimits=0.02,
        ylabel={Improvement (\%)},
        xticklabels = {
            $amazon0601$,$af\_shell6$,$apache2$,$atmosmodm$,$bcsstk32$,$cant$,$consph$,$Chevron3$,$Chevron4$,$crankseg\_1$,$crankseg\_2$,$CurlCurl\_3$,$eu-2005$,$hamrle3$,$hardesty1$,$human\_gene2$,$il2010$,$kim2$,$parabolic\_fem$,$pkustk04$,$pwtk$,$rgg\_n\_2\_20\_s0$,$rim$,$shar\_te2\-b3$,$Si87H76$,$test1$,$viscorocks$,$wiki-talk$,$x104$,$xenon2$, $GMean$
            },
        xticklabel style={rotate=90},
        minor x tick num = 1,
        minor tick length=0ex,
        xtick=data
        ]
\addplot[draw=black, fill=violet, error bars/.cd,y dir=both,y explicit] 
    table[x index=0,y index=3,y error plus index=2,y error minus index=1, col sep=comma] {results_perf_watt.csv};
    
\end{axis}
\end{tikzpicture}

\\

\end{tabular}
}

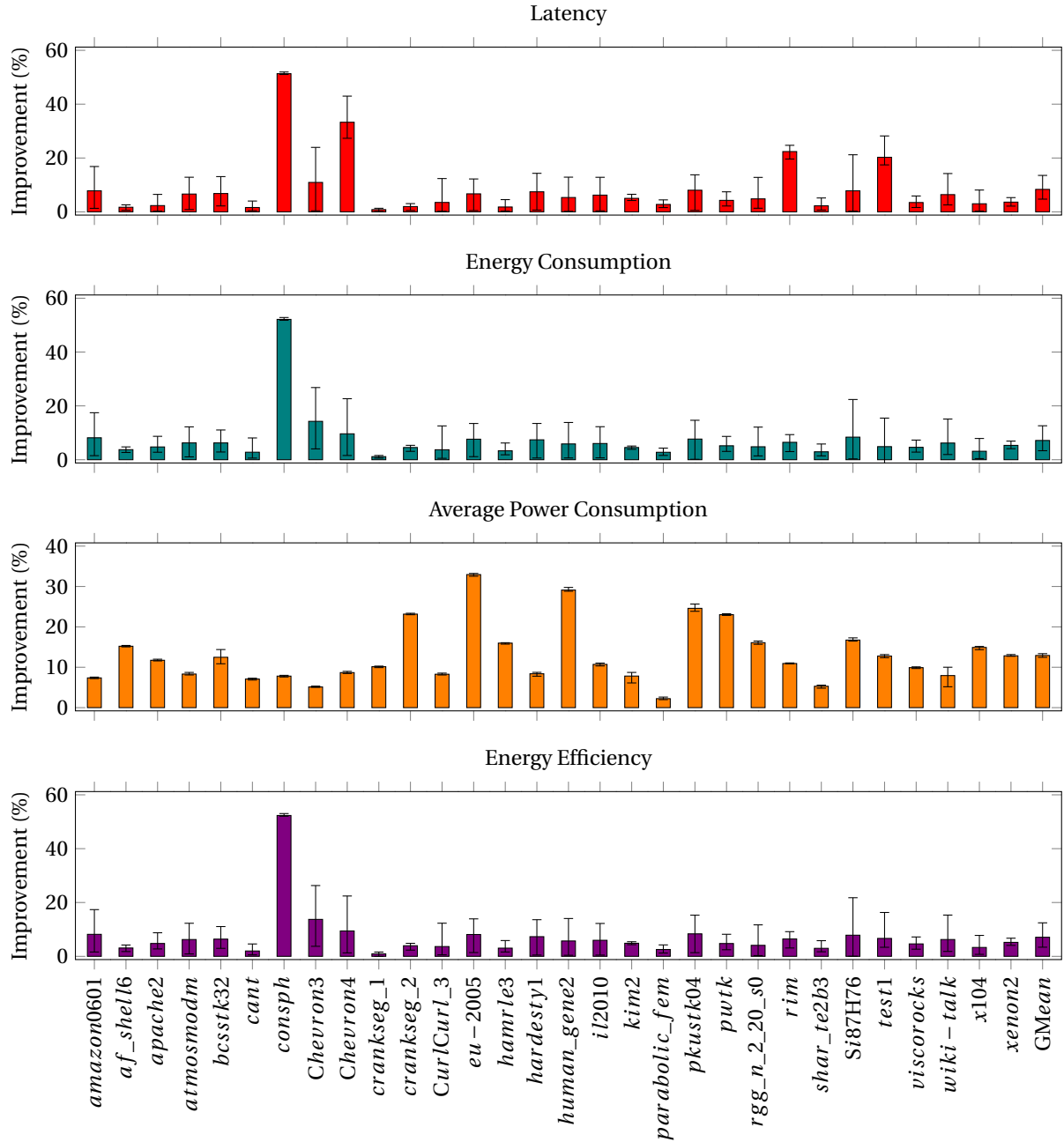
\captionof{figure}{Comparing Auto-SpMV with the default configuration settings for different Optimization objectives.}
\label{fig:result:offline}
\end{table*}

\subsection{Results of the \textit{run-time} Optimization Mode}
\label{sec:Results:Online}

Figure~\ref{fig:result:online} compares performance improvement of Auto-SpMV in the \textit{run-time} optimization mode over the default sparsity format, CSR, for different optimization objectives.
To ensure a fair comparison, we set the compilation parameters to the optimal values.
By optimizing the sparse format at run time, Auto-SpMV provides up to 34.6\% improvement in average power, as well as 99.7\% improvement in energy efficiency for all selected sparse matrices. 
As CSR is the best sparse format for the latency and energy consumption optimization objectives, Auto-SpMV's performance is identical to that of the default setting. 

\begin{table*}[htbp]
\centering
\resizebox{\textwidth}{!}{
\begin{tabular}{c}

\begin{tikzpicture}
\begin{axis}[ybar,
        title=Average Power Consumption,
        width=\textwidth,
        height=0.25\columnwidth,
        ymin=0,
        ymax=40,
        bar width=6pt,
        enlargelimits=0.02,
        ylabel={Improvement (\%)},
        xticklabels=empty,
        minor x tick num = 1,
        minor tick length=0ex
        ]
\addplot[draw=black, fill=orange] 
    table[x index=0,y index=1, col sep=comma] {online_results_avg_power.csv}; 
    
\end{axis}
\end{tikzpicture}
\\
\begin{tikzpicture}
\begin{axis}[ybar,
        title=Energy Efficiency,
        width=\textwidth,
        height=0.25\columnwidth,
        ymin=50,
        ymax=100,
        bar width=6pt,
        enlargelimits=0.02,
        ylabel={Improvement (\%)},
        xticklabels = {
            $amazon0601$,$af\_shell6$,$apache2$,$atmosmodm$,$bcsstk32$,$cant$,$consph$,$Chevron3$,$Chevron4$,$crankseg\_1$,$crankseg\_2$,$CurlCurl\_3$,$eu-2005$,$hamrle3$,$hardesty1$,$human\_gene2$,$il2010$,$kim2$,$parabolic\_fem$,$pkustk04$,$pwtk$,$rgg\_n\_2\_20\_s0$,$rim$,$shar\_te2\-b3$,$Si87H76$,$test1$,$viscorocks$,$wiki-talk$,$x104$,$xenon2$, $GMean$
            },
        xticklabel style={rotate=90},
        minor x tick num = 1,
        minor tick length=0ex,
        xtick=data
        ]
\addplot[draw=black, fill=violet] 
    table[x index=0,y index=1, col sep=comma] {online_results_perf_watt.csv};
    
\end{axis}
\end{tikzpicture}

\\

\end{tabular}
}
\captionof{figure}{Auto-SpMV run time optimization over the default sparse format for different Optimization objectives.}
\label{fig:result:online}
\end{table*}

\subsection{Classification Results}
\label{sec:results:classification}

Table~\ref{tab:results:classification} shows the best classification results achieved by Auto-SpMV for predicting the best configuration settings.
We report accuracy and F1-Score as the evaluation metrics.
Results show that Auto-SpMV successfully devises highly accurate learning models with 100\% accuracy for predicting the optimal settings for TB size, \texttt{maxrregcount}, and memory hierarchy configuration.  
Note that the best result is provided by tuning the hyperparameters of the decision tree classifier \cite{safavian1991survey}.

\begin{table}[htbp]
    \centering
    \caption{The classification accuracy of the optimal learning models (Decision Tree) used for predicting the best configuration settings for running SpMV kernels on GPU.}
    \resizebox{\columnwidth}{!}{
    \begin{tabular}{|c|c|c|c|c|c|c|c|c|}
    \hline
        \textbf{Classification} & \multicolumn{2}{c|}{\textbf{Latency}}  & \multicolumn{2}{c|}{\textbf{Energy}}  & \multicolumn{2}{c|}{\textbf{Average~Power}} & \multicolumn{2}{c|}{\textbf{Energy~Efficiency}}   \\ \cline{2-9} 
       (\%) & Acc. & F1-Score & Acc.  & F1-Score  & Acc.  & F1-Score  & Acc.  & F1-Score   \\ \hline
        TB Size  & 100 & 100 & 100 & 70 & 100 & 87.5 & 100 & 100 \\ \hline
        \texttt{maxregcount} & 100 & 100 & 100 & 88.9 & 100 & 92.5 & 100 & 92.9 \\ \hline
        Memory & 100 & 91.7 & 100 & 81.8 & 100 & 50 & 100 & 91.7 \\ \hline
    \end{tabular}}
    \label{tab:results:classification}
\end{table}
Table~\ref{tab:results:classification:comparison} compares the classification results of Auto-SpMV with state-of-the-art. 
At least 10\% more accurate classification is achieved by Auto-SpMV compared to the state-of-the-art.
By evaluating different learning models and constructing a large-scale optimization space, Auto-SpMV is able to learn efficiently.

\begin{table}[htbp]
\centering
\caption{Comparison of classification results of Auto-SpMV with state-of-the-art. }
\label{tab:results:classification:comparison}
\resizebox{\textwidth}{!}{%
\begin{tabular}{ccccc}
\hline
\multicolumn{1}{c}{} &
  \textbf{Learning Model} &
  \textbf{Hardware Architecture} &
  \textbf{Accuracy (Execution Time)} &
  \textbf{Accuracy (Energy)} \\ \hline
BestSF \cite{[31]}  & SVM           & Maxwell & 82\%  & -    \\
\cite{dufrechou2021selecting} &
  Bagged Trees Classifier & Pascal & 89\% & 84\% \\
\cite{zhao2018bridging} & CNN           & Pascal  & 90\%  & -   \\  \hline
Auto-SpMV (ours) & Decision Tree & Turing  & \textbf{100\%} & \textbf{100\%} \\ \hline
\end{tabular}%
}
\end{table}

\subsection{Regression Results}
\label{sec:Results:Prediction:Regression}

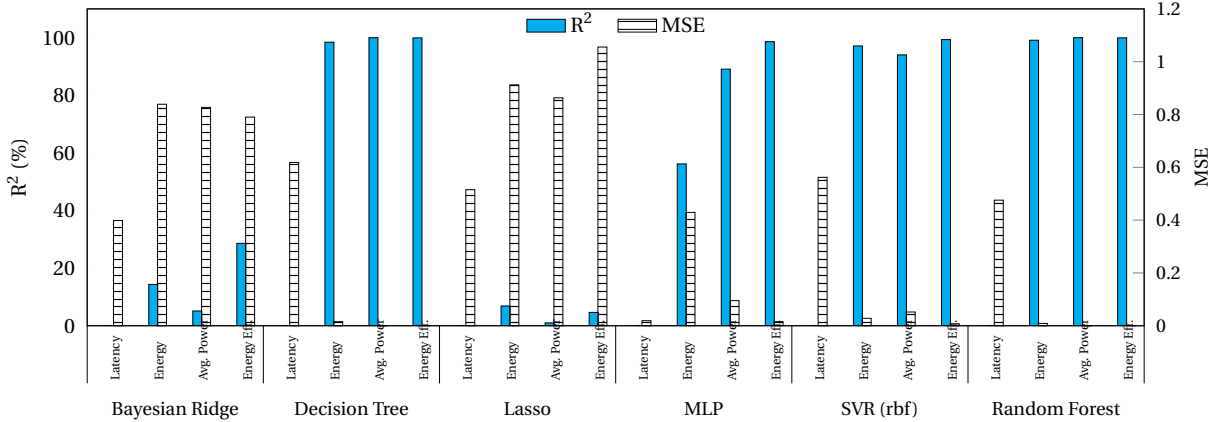
\begin{figure*}[ht]
\centering
\resizebox{\linewidth}{!}{
    \pgfmathsetmacro{\BarOffset}{-0.001}

    \pgfplotsset{
    MyAxis/.style={
            ybar,
            scale only axis,
            width=\textwidth,
            height=0.3\columnwidth,
            ymin=0,
            xmin=-0.5,
            enlarge x limits=0.00,
            xmax=23.5, 
            area legend,
            bar width=4pt
      }
    }
{
    \begin{tikzpicture}
    \begin{axis}[
        ylabel={R\textsuperscript{2} (\%)},
        ymax=110,
        xtick={0,...,23},
        xtick pos=left,
        ytick pos=left,
        xticklabels = {Latency, Energy, Avg. Power, Energy Eff., Latency, Energy, Avg. Power, Energy Eff., Latency, Energy, Avg. Power,  Energy Eff., Latency, Energy, Avg. Power,  Energy Eff., Latency, Energy, Avg. Power,  Energy Eff., Latency, Energy, Avg. Power,  Energy Eff. },
        xticklabel style={rotate=90, font=\tiny, below right=-0.4em, xshift=-6.8ex},
        ymin=0,
        yticklabel style={xshift=-0.5ex},
        tickwidth=5pt,
        extra x ticks={-0.5,3.5,7.5,...,23.5},
        extra x tick labels={},
        tickwidth=0mm,
        every node near coord/.append style={anchor=north},
        xtick style={draw=black},
        extra x tick style={tickwidth=0.9cm},
        minor x tick style = {opacity=1},
        MyAxis
    ]
    \addplot[draw=black,fill=cyan] table[x expr=\coordindex,y index=1] \dataset;
    \label{R2plot}
    \end{axis}

    \begin{axis}[
        name=ax2,
        ymax=1.2,
        MyAxis,
        font=\small,
        ylabel={MSE},
        ytick pos=right,
        yticklabel style={xshift=0.5ex},
        clip=false,
        xtick={1.5,5.5,9.5,13.5,17.5,21.5},
        xticklabels={Bayesian~Ridge, Decision~Tree, Lasso, MLP, SVR (rbf), Random~Forest},
        xticklabel style={yshift=-9mm},
        tickwidth=5pt,
        xtick style={draw=none}
    ]
    \addplot[draw=black,fill=black!40, pattern=horizontal lines, pattern color=black] table[x expr=\coordindex+0.2\BarOffset,y index=2] \dataset; 
    \label{MSEplot}
    \end{axis}

    \node [below=-5.0cm] at (ax2.south) {\ref{R2plot} R\textsuperscript{2} \quad  \ref{MSEplot} MSE};
    \end{tikzpicture}
    }
  }
 \captionof{figure}{Evaluating different regression models for the estimation of different optimization objectives.}
  \label{Fig:regression}
\end{figure*}

Figure~\ref{Fig:regression} shows the regression results of estimating different optimization objectives.
We report R\textsuperscript{2} correlation, and MSE estimation metrics to identify the optimal learning model for each Optimization objective. 
Results show that random forest is the best learning model for estimating the energy consumption and energy efficiency of an SpMV kernel by providing 99.11\% and 99.94\% R\textsuperscript{2} correlation, and 0.0086 and 0.00063 MSE, respectively.
The decision tree is the best learning model for estimating average power by providing R\textsuperscript{2} = 99.99\% and MSE = 2.82$\times10^{-5}$.  
MLP is the best learning model for estimating latency by providing MSE = 1.9$\times10^{-2}$.
Our success in using ML algorithms for estimating different objective functions is mainly due to 1) using a large training dataset that covers a broad range of sparse matrices with different sizes and sparsity features, and 2) tweaking the hyperparameters of learning models.

\pgfplotstableread{
0 0 0.39839
1 14.350166 0.838839969
2 5.1236705 0.827119641
3 28.6002203 0.79016061
4 0 0.6183151
5 98.4235176 0.015439802
6 99.996768 2.82E-05
7 99.933256 0.000738637
8 0 0.5152626
9 6.8561277 0.912235311
10 0.9846399 0.863203178
12 4.6308455 1.055422719
13 0 0.01911
14 56.1819457 0.429146603
15 89.085611 0.09515024
16 98.6127782 0.01535198
17 0 0.5621047
18 97.1299248 0.02810903
19 94.0285486 0.052058345
20 99.3295503 0.007419672
21 0 0.4756270
22 99.1155715 0.008661942
23 99.997283 2.37E-05
24 99.9435177 0.000625073
}\dataset

\subsection{Run Time Prediction Overhead}
\label{sec:Results:Overhead}

In this section, we analyze the overhead of the optimization modes. 
Due to leveraging an overhead prediction mechanism (Section~\ref{sec:Method:Online}), Auto-SpMV does not perform unnecessary conversions when the conversion overhead is higher than the optimization benefit.
The overhead of the optimization modes is computed based on the f\_latency + o\_latency + p\_latency + c\_latency.
f\_latency is the elapsed time to extract features, o\_latency is the time needed to predict the conversion overhead, p\_latency is the time needed to predict the optimal sparse format, and c\_latency is the latency of converting to the predicted sparse format. 
The input matrix is often saved in a default sparse format, which is the COO sparse format in SuiteSparse \cite{davis2011university}.
Note that all the steps of the run time decision-making phase are executed on the CPU side using parallel implementations with Numpy Python arrays and matrices programming library. 

Our measurements show that overall run time overhead is dominated by the sum of f\_latency and c\_latency since the o\_latency and p\_latency are constant values between $\approx$20 milliseconds while two other latencies are introduced according to the characteristics of the given input matrix. 
Table~\ref{tab:overhead} shows the overhead of selected matrices if we consider lower latency as the optimization objective.
The average f\_latency+c\_latency of our dataset is 24.2 seconds for the dataset while the maximum value is 87.8 seconds for a sparse matrix with nearly $1.9\times10^{7}$ non-zero elements ($nnz$).
As also shown in \cite{benatia2016machine, [31]}, this overhead is negligible for applications running large-scale sparse matrices such as Preconditioned Conjugate Gradient method which is an iterative solver. 
Thus, the total prediction overhead is insignificant for applications where an SpMV kernel is executed several times with the same input matrix in iterative methods such as eigenvalue problems and linear systems \cite{li2019eigenvalues, [31]}.

\begin{table}[htbp]
\centering
\caption{Optimization Overhead of the Auto-SpMV Optimization Modes in seconds. Sorted based on the ascending order of $nnz$.}
\label{tab:overhead}
\resizebox{\columnwidth}{!}{%
\begin{tabular}{ccccc}
\hline
\textbf{Matrix Name} & \textbf{$nnz$} & \textbf{f\_latency} & \textbf{c\_latency} & \textbf{f\_latency+c\_latency} \\\hline
shar\_te2-b3       & 800800   & 1.71875  & 1.625    & 3.34375  \\ 
rim                & 1014951  & 1.578125 & 2.046875 & 3.625    \\
bcsstk32           & 1029655  & 1.71     & 2.125    & 3.835    \\
il2010             & 1082232  & 3.625    & 2.5      & 6.125    \\
viscorocks         & 1162244  & 1.90625  & 2.4375   & 4.34375  \\
cant               & 2034917  & 3.4531   & 4.59     & 8.0431   \\
parabolic\_fem     & 2100225  & 5.46875  & 4.984375 & 10.45313 \\
pkustk04           & 2137125  & 3.78     & 4.53125  & 8.31125  \\
apache2            & 2766523  & 7.84     & 6.06     & 13.9     \\
consph             & 3046907  & 5.375    & 6.65625  & 12.03125 \\
wiki-talk-temporal & 3309592  & 10.4375  & 7.3281   & 17.7656  \\
amazon0601         & 3387388  & 7.125    & 7.171875 & 14.29688 \\
Chevron3           & 3413113  & 7.0625   & 7.328125 & 14.39063 \\
xenon2             & 3866688  & 6.75     & 9.375    & 16.125   \\
x104               & 5138004  & 9.093    & 11.76563 & 20.85863 \\
crankseg\_1        & 5333507  & 9.78025  & 11.75    & 21.53025 \\
Si87H76            & 5451000  & 9.828125 & 11.90625 & 21.73438 \\
Hamrle3            & 5514242  & 15.09375 & 12.89063 & 27.98438 \\
pwtk               & 5926171  & 11.39    & 13.8593  & 25.2493  \\
Chevron4           & 6376412  & 13.76563 & 14.71875 & 28.48438 \\
Hardesty1          & 6539157  & 15.09375 & 14.5625  & 29.65625 \\
rgg\_n\_2\_20\_s0    & 6891620      & 15.32813            & 15.34375            & 30.67188                       \\
crankseg\_2        & 7106348  & 13.03125 & 15.25    & 28.28125 \\
CurlCurl\_3        & 7382096  & 17.89063 & 18.8125  & 36.70313 \\
human\_gene2       & 9041364  & 17.01563 & 21.70313 & 38.71875 \\
af\_shell6         & 9046865  & 18.78125 & 21.4687  & 40.24995 \\
atmosmodm          & 10319760 & 24.14063 & 23.90625 & 48.04688 \\
kim2               & 11330020 & 22.70313 & 27.10938 & 49.8125  \\
test1              & 12968200 & 23.78125 & 30.03125 & 53.8125  \\
eu-2005            & 19235140 & 39.82813 & 47.98438 & 87.8125 \\ \hline
\end{tabular}%
}
\end{table}

\subsection{GPU Sensitivity of Learning Models}
\label{sec:Results:GPU-Sensitivity}

We must verify that our classification results are independent of the underneath GPU architecture. 
To this end, we repeat a subset of evaluations for $amazon0601$, $crankseg\_2$, $bcsstk32$, $x104$, $il2010$, $Chevron3$ sparse matrices on the NVIDIA\textsuperscript{®} 1080 GPU device with the Pascal architecture.
Then, we compare the performance of predicted configuration settings provided by the Auto-SpMV classifier with the actual measurements on the NVIDIA\textsuperscript{®} 1080 GPU device. 
Note that the prediction results of Auto-SpMV are based on running SpMV kernels on the NVIDIA\textsuperscript{®} GTX 1650-mobile device with the Turing architecture.
Figure~\ref{fig:GPU_MUlti} compares the optimal measurements of running SpMV kernels on an NVIDIA\textsuperscript{®} GPU with the results predicted by Auto-SpMV.
Results are normalized to the actual measurements. 
Results show that Auto-SpMV can predict the optimal configuration settings (TB Size and \texttt{maxrregcount}) with a marginal different (up to 2\% performance loss).
Accordingly, Auto-SpMV predicts the optimal configuration settings without relying on the underneath GPU architecture.

\begin{table}[htbp]
\centering
\resizebox{\columnwidth}{!}{
\begin{tabular}{cc}

\multicolumn{2}{c}{
\begin{tikzpicture}
\begin{customlegend}[legend columns=4,legend style={text opacity = 1,row sep=0pt, font=\fontsize{24}{8}\selectfont, column sep=2ex},
        legend entries={{Actual Latency},
                        {Predicted Latency},
                        {Actual Energy},
                        {Predicted Energy},
                        {Actual Avg.~Power},
                        {Predicted Avg.~Power},
                        {Actual Energy~Eff.},
                        {Predicted Energy~Eff.}
                        }]
        \addlegendimage{mark=square*, mark size=7pt, only marks, red, thick, draw=black}
        \addlegendimage{ mark=square*, mark size=7pt, only marks, pattern=north east lines, pattern color= red, thick, draw=black}
        \addlegendimage{mark=square*, mark size=7pt, only marks, teal, thick, draw=black}
        \addlegendimage{mark=square*, mark size=7pt, only marks, pattern=north east lines, pattern color= teal, thick, draw=black}
        \addlegendimage{mark=square*, mark size=7pt, only marks, blue, thick, draw=black}
        \addlegendimage{mark=square*, mark size=7pt, only marks, pattern=north east lines, pattern color= blue, thick, draw=black}
        \addlegendimage{mark=square*, mark size=7pt, only marks, violet, thick, draw=black}
        \addlegendimage{mark=square*, mark size=7pt, only marks, pattern=north east lines, pattern color= violet, thick, draw=black}
        \end{customlegend}
\end{tikzpicture}}
\\
\begin{tikzpicture}
\begin{axis}[
    title={\Large \textbf{amazon0601}},
    ybar=8pt,
    width=\columnwidth,
    height=0.4\columnwidth,
    font=\Large,
    enlargelimits=0.44,
    grid=major, 
    grid style={dashed,gray!90}, 
    ylabel={\Large \textbf{Norm. Measurement}},
    symbolic x coords={TB Size, \texttt{maxregcount}},
    xtick=data,
    bar width=10pt,
    nodes near coords align={vertical},
    ]
    
\addplot [red, fill, draw=black ] coordinates {(TB Size,1) (\texttt{maxregcount},1)};
\addplot [postaction={pattern=north east lines}, black, pattern color=red] coordinates {(TB Size,1) (\texttt{maxregcount},1)};
\addplot [teal, fill, draw=black] coordinates {(TB Size,1) (\texttt{maxregcount},1) };
\addplot [postaction={pattern=north east lines}, black, pattern color=teal] coordinates {(TB Size,1) (\texttt{maxregcount},1) };
\addplot  [blue, fill, draw=black] coordinates {(TB Size,1) (\texttt{maxregcount},1) };
\addplot  [postaction={pattern=north east lines}, black, pattern color=blue] coordinates {(TB Size,1) (\texttt{maxregcount},0.98) };
\addplot [violet, fill, draw=black] coordinates {(TB Size,1) (\texttt{maxregcount},1) };
\addplot [postaction={pattern=north east lines}, black, pattern color=violet] coordinates {(TB Size,1) (\texttt{maxregcount},1) };
\end{axis}
\end{tikzpicture}

&

\begin{tikzpicture}
\begin{axis}[
    title={\Large \textbf{crankseg\_2}},
    ybar=8pt,
    width=\columnwidth,
    height=0.4\columnwidth,
    font=\Large,
    enlargelimits=0.44,
    grid=major, 
    grid style={dashed,gray!90}, 
    ylabel={\Large \textbf{Norm. Measurement}},
    symbolic x coords={TB Size, \texttt{maxregcount}},
    xtick=data,
    bar width=10pt,
    nodes near coords align={vertical},
    ]
    
\addplot [red, fill, draw=black ] coordinates {(TB Size,1) (\texttt{maxregcount},1)};
\addplot [postaction={pattern=north east lines}, black, pattern color=red ] coordinates {(TB Size,1) (\texttt{maxregcount},1)};
\addplot [teal, fill, draw=black] coordinates {(TB Size,1) (\texttt{maxregcount},1) };
\addplot [postaction={pattern=north east lines}, black, pattern color=teal] coordinates {(TB Size,1) (\texttt{maxregcount},1) };
\addplot  [blue, fill, draw=black] coordinates {(TB Size,1) (\texttt{maxregcount},1) };
\addplot  [postaction={pattern=north east lines}, black, pattern color=blue] coordinates {(TB Size,1) (\texttt{maxregcount},0.99) };
\addplot [violet, fill, draw=black] coordinates {(TB Size,1) (\texttt{maxregcount},1) };
\addplot [postaction={pattern=north east lines}, black, pattern color=violet] coordinates {(TB Size,1) (\texttt{maxregcount},1) };

\end{axis}
\end{tikzpicture}

\\

\begin{tikzpicture}
\begin{axis}[
    title={\Large \textbf{bcsstk32}},
    ybar=8pt,
    width=\columnwidth,
    height=0.4\columnwidth,
    font=\Large,
    enlargelimits=0.44,
    grid=major, 
    grid style={dashed,gray!90}, 
    ylabel={\Large \textbf{Norm. Measurement}},
    symbolic x coords={TB Size, \texttt{maxregcount}},
    xtick=data,
    bar width=10pt,
    nodes near coords align={vertical},
    ]
    
\addplot [red, fill, draw=black ] coordinates {(TB Size,1) (\texttt{maxregcount},1)};
\addplot [postaction={pattern=north east lines}, black,  pattern color=red] coordinates {(TB Size,1) (\texttt{maxregcount},1)};
\addplot [teal, fill, draw=black] coordinates {(TB Size,1) (\texttt{maxregcount},1) };
\addplot [postaction={pattern=north east lines}, black,  pattern color=teal] coordinates {(TB Size,1) (\texttt{maxregcount},0.99) };
\addplot  [blue, fill, draw=black] coordinates {(TB Size,1) (\texttt{maxregcount},1) };
\addplot  [postaction={pattern=north east lines}, black,  pattern color=blue] coordinates {(TB Size,1) (\texttt{maxregcount},0.99) };
\addplot [violet, fill, draw=black] coordinates {(TB Size,1) (\texttt{maxregcount},1) };
\addplot [postaction={pattern=north east lines}, black,  pattern color=violet] coordinates {(TB Size,1) (\texttt{maxregcount},1) };

\end{axis}
\end{tikzpicture}

&

\begin{tikzpicture}
\begin{axis}[
    title={\Large \textbf{x104}},
    ybar=8pt,
    width=\columnwidth,
    height=0.4\columnwidth,
    font=\Large,
    enlargelimits=0.44,
    grid=major, 
    grid style={dashed,gray!90}, 
    ylabel={\Large \textbf{Norm. Measurement}},
    symbolic x coords={TB Size, \texttt{maxregcount}},
    xtick=data,
    bar width=10pt,
    nodes near coords align={vertical},
    ]
    
\addplot [red, fill, draw=black ] coordinates {(TB Size,1) (\texttt{maxregcount},1)};
\addplot [postaction={pattern=north east lines}, black,  pattern color=red ] coordinates {(TB Size,1) (\texttt{maxregcount},1)};
\addplot [teal, fill, draw=black] coordinates {(TB Size,1) (\texttt{maxregcount},1) };
\addplot [postaction={pattern=north east lines}, black,  pattern color=teal] coordinates {(TB Size,1) (\texttt{maxregcount},1) };
\addplot  [blue, fill, draw=black] coordinates {(TB Size,1) (\texttt{maxregcount},1) };
\addplot  [postaction={pattern=north east lines}, black,  pattern color=blue] coordinates {(TB Size,1) (\texttt{maxregcount},1) };
\addplot [violet, fill, draw=black] coordinates {(TB Size,1) (\texttt{maxregcount},1) };
\addplot [postaction={pattern=north east lines}, black,  pattern color=violet] coordinates {(TB Size,1) (\texttt{maxregcount},1) };

\end{axis}
\end{tikzpicture}

\\

\begin{tikzpicture}
\begin{axis}[
    title={\Large \textbf{il2010}},
    ybar=8pt,
    width=\columnwidth,
    height=0.4\columnwidth,
    font=\Large,
    enlargelimits=0.44,
    grid=major, 
    grid style={dashed,gray!90}, 
    ylabel={\Large \textbf{Norm. Measurement}},
    symbolic x coords={TB Size, \texttt{maxregcount}},
    xtick=data,
    bar width=10pt,
    nodes near coords align={vertical},
    ]
    
\addplot [red, fill, draw=black ] coordinates {(TB Size,1) (\texttt{maxregcount},1)};
\addplot [postaction={pattern=north east lines}, black,  pattern color=red ] coordinates {(TB Size,1) (\texttt{maxregcount},1)};
\addplot [teal, fill, draw=black] coordinates {(TB Size,1) (\texttt{maxregcount},1) };
\addplot [postaction={pattern=north east lines}, black,  pattern color=teal] coordinates {(TB Size,1) (\texttt{maxregcount},1) };
\addplot  [blue, fill, draw=black] coordinates {(TB Size,1) (\texttt{maxregcount},1) };
\addplot  [postaction={pattern=north east lines}, black,  pattern color=blue] coordinates {(TB Size,1) (\texttt{maxregcount},0.99) };
\addplot [violet, fill, draw=black] coordinates {(TB Size,1) (\texttt{maxregcount},1) };
\addplot [postaction={pattern=north east lines}, black,  pattern color=violet] coordinates {(TB Size,1) (\texttt{maxregcount},1) };

\end{axis}
\end{tikzpicture}

&

\begin{tikzpicture}
\begin{axis}[
    title={\Large \textbf{Chevron3}},
    ybar=8pt,
    width=\columnwidth,
    height=0.4\columnwidth,
    font=\Large,
    enlargelimits=0.44,
    grid=major, 
    grid style={dashed,gray!90}, 
    ylabel={\Large \textbf{Norm. Measurement}},
    symbolic x coords={TB Size, \texttt{maxregcount}},
    xtick=data,
    bar width=10pt,
    nodes near coords align={vertical},
    ]
    
\addplot [red, fill, draw=black ] coordinates {(TB Size,1) (\texttt{maxregcount},1)};
\addplot [postaction={pattern=north east lines}, black,  pattern color=red ] coordinates {(TB Size,1) (\texttt{maxregcount},1)};
\addplot [teal, fill, draw=black] coordinates {(TB Size,1) (\texttt{maxregcount},1) };
\addplot [postaction={pattern=north east lines}, black,  pattern color=teal] coordinates {(TB Size,1) (\texttt{maxregcount},1) };
\addplot  [blue, fill, draw=black] coordinates {(TB Size,1) (\texttt{maxregcount},1) };
\addplot  [postaction={pattern=north east lines}, black,  pattern color=blue] coordinates {(TB Size,1) (\texttt{maxregcount},0.98) };
\addplot [violet, fill, draw=black] coordinates {(TB Size,1) (\texttt{maxregcount},1) };
\addplot [postaction={pattern=north east lines}, black,  pattern color=violet] coordinates {(TB Size,1) (\texttt{maxregcount},1) };

\end{axis}
\end{tikzpicture}

\end{tabular}
}

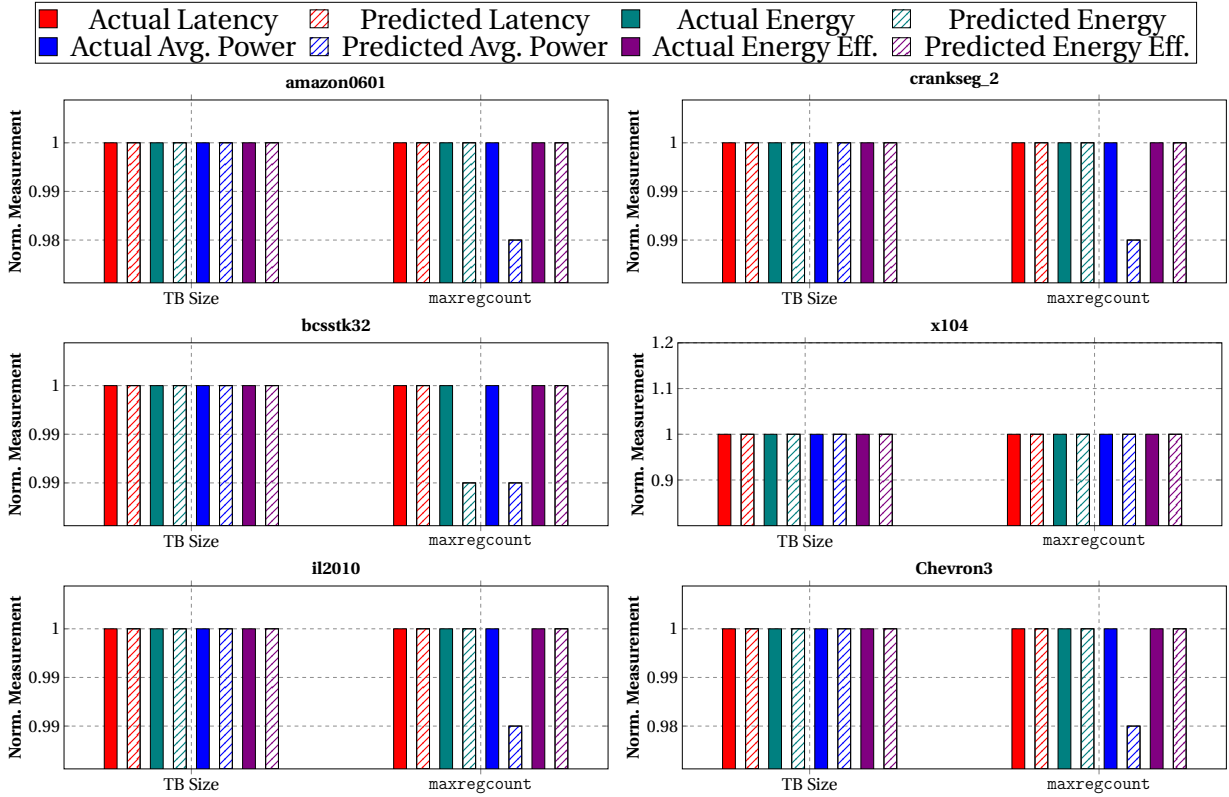
\captionof{figure}{Evaluating efficiency of Auto-SpMV for predicting optimal configuration settings of new GPU devices.}
\label{fig:GPU_MUlti}
\end{table}

\section{Discussion}
\label{sec:Discussion}

Our findings are summarized as follows:

\begin{enumerate}
    \item There are no optimal configuration settings for all the sparse input matrices even on the same sparse format.
    
    \item Carelessly selecting compiler parameters and storage format  results in a significant performance loss of 13.3\%, 5.4\%, 2.0\%, and 13.9\% on average for latency, energy consumption, average power, and energy efficiency optimization objective, respectively (Figure~\ref{fig:result:offline}). 
    
    \item Considering the CSR as the default sparse format, a careless choice of sparse format (\textit{compile-time} mode) leads to a significant performance loss of 13.4\% and 86.89\% on average for the average power and energy efficiency optimization objectives, respectively (Figure~\ref{fig:result:online}).
    
    \item Programmers cannot achieve the maximum improvement even when they try to select the optimal thread block sizes since \texttt{maxrregcount} and memory hierarchy configuration are selected by the CUDA\textsuperscript{®} compiler.
    
    \item Results demonstrate that CSR is not the best format for average power consumption and energy efficiency optimization objectives. 
    
    \item Increasing \texttt{maxrregcount} can reduce memory latency. In contrast, reducing \texttt{maxrregcount} can increase occupancy by allowing more threads to be active and thereby more warps to run. Therefore, an inefficient \texttt{maxrregcount} leads to  1) register spilling in global memory, and 2) decreased occupancy rate, resulting in increased latency and significant energy consumption \cite{zardoshti2016adaptive}. 

    \item Increasing the thread block size can result in more threads per SM, and therefore, a higher thread occupancy. On the other hand, when thread blocks are suspended, fewer thread blocks are active, resulting in less parallelism. Therefore, programmers should carefully consider the trade-off between increasing the occupancy of SMs and decreasing the number of thread blocks in order to achieve a higher level of parallelism, and as a result, be more efficient in terms of energy consumption and performance.
    
    \item In order to fully utilize GPU computing power, memory access efficiency is essential. By determining the optimal size of the L1 cache and shared memory based on the sparse input matrix features, we can ensure better memory access efficiency and thus improved performance and energy efficiency.
    
    \item The selection of a sparse format has a profound impact on data locality, cache performance, and memory bandwidth utilization. Results show that a significant amount of processing time and energy can be saved by automating the selection of the sparse format of SpMV kernels on GPU. 
    
\end{enumerate}

\section{Related Work}
\label{sec:RelatedWork}

To the best of our knowledge, Auto-SpMV is the first automated framework that optimizes the performance and energy efficiency of SpMV kernels on GPU considering hardware and software optimization simultaneously. It is also the first study to demonstrate that compiler parameters play an integral role in GPU utilization, run time performance, and energy consumption. The efficient running of SpMV kernels on GPU has been extensively researched in the past. Prior studies are categorized as 1) proposing novel SpMV sparse formats, 2) software optimization techniques for SpMV kernels, and 3) performance or energy estimation of SpMV kernels on GPU. In this section, we briefly discuss these methods and compare them with this paper.

\textbf{SpMV sparse formats.} \cite{[2], [3]} present the earliest studies on accelerating SpMV kernels on GPU. They provide a detailed analysis of the memory access pattern of CUDA\textsuperscript{®} implementation of the classical sparse formats, including ELL, CSR, and COO. During the last decade, many attempts have been made to propose a range of SpMV sparse formats \cite{[5], [6], [8], [9], [10], [11], langr2015evaluation}. In general, these formats significantly improve memory saving and can be applied to any sparse matrix. However, as we demonstrate in Section~\ref{sec:Motivation}, carelessly selecting the sparse format and compiler parameters incurs significant performance degradation as a result of low GPU utilization. To maximize the performance of SpMV kernels on GPU, in this paper, we predict the best configuration settings using highly optimized machine learning models trained on a large-scale dataset.

Several SpMV sparse formats utilized specific sparse matrix characteristics in order to provide higher efficient storage usage and indexing \cite{kourtis2008optimizing, kourtis2011csx, li2013smat}. For example, CSR-VI and CSR-DU use non-zero values similarity to achieve more efficient sparse matrix compression \cite{kourtis2008optimizing}. CSX first analyzes a sparse matrix to find patterns such as repeated values or dense diagonals. Then encodes the matrix using sparse formats appropriate for the founded patterns \cite{kourtis2011csx}. However, these approaches are limited in applicability to matrices containing specific sparsity features, besides, requiring costly pre-processing to identify sparsity features. Therefore, these specific formats are neglected in our work. 

\textbf{Software optimization techniques for SpMV kernels.}
Many existing works investigate the potential of different performance enhancement techniques such as auto-tuning \cite{[12],hou2017auto}, partitioning the sparse matrix and combining sparse formats \cite{[17],[19]}, improving the reuse of cache and registers \cite{[20],[21]} and hardware-software co-design optimizations \cite{[22],[24]}. Despite the performance improvement of these techniques, balancing the trade-off between energy consumption and performance in running SpMV on energy-hungry GPUs has not been addressed in the literature. 

\textbf{Performance or energy estimation of SpMV kernels on GPU.} 
Many studies focused on accelerating SpMV kernels by proposing a performance estimation model to select the best configuration settings \cite{[13], [14], [15], [16], li2020adaptive, zhao2018bridging, benatia2016machine, [31]}. These techniques mainly leverage machine learning techniques to learn the features of the sparse matrix and provide highly accurate predictions. In contrast with the state-of-the-art \cite{li2020adaptive}, our work considers more sparse matrix features to improve prediction accuracy. Besides latency and energy consumption \cite{ [27], [29], [30]}, we model the energy efficiency and average power consumption of SpMV kernels to propose the best configuration settings for low-power applications, e.g., running a geometric flow task on embedded devices \cite{wang2021tc}. We also optimize different learning models using an AutoML tool to find the best-performing prediction model for a given sparse matrix.

\section{Conclusion}
\label{sec:Conclusion}

This paper proposes Auto-SpMV, an automated framework for optimizing SpMV kernels on GPU by predicting the best-performing sparse format and compiler parameters.
Auto-SpMV enables optimizing different objectives including latency, energy consumption, average power consumption, and energy efficiency.
Our experimental results using real-world benchmarks reveal that Auto-SpMV achieves remarkable performance and energy efficiency compared to the best default compiler parameters and sparse format.
We hope our new findings on the performance characteristics of SpMV kernels on GPU inspire new techniques for improving the performance of future performance modeling algorithms.

\end{document}